\documentclass[%
 reprint,
nofootinbib,
 amsmath,amssymb,
 aps,
]{revtex4-2}

\usepackage{graphicx}
\usepackage{amsfonts,amssymb,amstext,amsmath,amsthm,verbatim,times,cancel,epsfig,bbold}
\usepackage{hyperref}
\usepackage[english]{babel}
\usepackage{xcolor}
\usepackage{enumitem}
\usepackage{dcolumn}
\usepackage{bm}

\begin{document}

\preprint{APS/123-QED}

\title{Gravitational waves away from a Schwarzschild black hole: full spectrum analysis}
\author{Marco Refuto}
\email{marco.refuto@unina.it}
\affiliation{Dipartimento di Fisica Ettore Pancini, Università di Napoli “Federico II”,
Complesso Univ.\ Monte S.\ Angelo, I-80126 Napoli, Italy}
\affiliation{INFN, Sezione di Napoli, Complesso Univ.\ Monte S.\ Angelo, I-80126 Napoli, Italy}
\author{Giampiero Esposito}
\email{giampiero.esposito@na.infn.it}
\affiliation{Dipartimento di Fisica Ettore Pancini, Università di Napoli “Federico II”,
Complesso Univ.\ Monte S.\ Angelo, I-80126 Napoli, Italy}
\affiliation{INFN, Sezione di Napoli, Complesso Univ.\ Monte S.\ Angelo, I-80126 Napoli, Italy}

\begin{abstract}
Gravitational waves as one approaches Schwarzschild null infinity are the core of our 
investigation, treated as the spin-weighted generalisation of a massless scalar 
field. In particular, we compute their full spectrum, consistent with 
observations coming from LIGO and VIRGO experiments. This paper continues 
recent work by the authors, where the Poincar\'e asymptotic expansion was exploited 
in black hole perturbation theory. The mathematical formalism we developed 
is now intimately connected with gravitational wave phenomenology, providing 
new mathematical tools to investigate their features.
\end{abstract}

\maketitle

\section{Introduction}

Natural phenomena are described by mathematics, in a long-standing effort aimed at  
understanding and predicting the observations of what happens around us. However, 
history tells us that such a process is not linear at all. There are certain research 
lines where experiments came before a theory able to described them, and a new mathematics 
was required in order to predicts data coming from, e.g., the spectroscopy of atoms 
and molecules. The early days of Quantum Mechanics provide one of the most known examples 
of this. On the other hand, the theoretical framework goes far beyond the technical 
possibilities, and we have to wait a long time in order to have experimental confirmations 
of our mathematical predictions. Gravitational waves, 
discovered a century after Einstein's theory predicted them, 
fall exactly in this category, paving the way for 
completely new possibilities in such a research avenue. 
The scientific community needed almost forty years in order to prove an existence
and uniqueness theorem for solutions of the vacuum Einstein equations in the
de Donder gauge with nonanalytic initial data
(cf. Ref. \cite{FoursBruhat1952ThormeDP}).

Our efforts aim at developing 
a new mathematical tool able to compute with the desired level of accuracy the behavior of 
black hole perturbations in the asymptotic spacetime region. The Poincar\'e asymptotic 
series (cf. Refs \cite{Poincare1, Poincare2}) can be used to solve the radial 
wave equation for large values of the radial coordinate. One of the most useful advantages 
is given by the possibility to go beyond the leading order often considered in the literature, 
arriving at highly precise analytic computations. In our first work (cf. Ref. \cite{Esposito:2025jsq}), 
we introduced the method in the easiest scenario: a scalar field in Schwarzschild 
spacetime. Our second investigation (cf. Ref. \cite{Esposito:2026las}) concerns the 
extension of such a mathematical framework for every Petrov D spacetime, since the 
existence of the so-called principal tensor allows us to separate the equation describing 
the dynamics of spin-weighted fields and therefore to consider just the radial part, 
which is nothing but a confluent Heun equation. Once the mathematics is understood, the 
natural continuation is finding a connection with experimental data, in order to ensure the 
validity of the method and the possibility to predict beyond-the-leading-order effects, which 
are often out of control, even if of central importance for high energy physics. One of the 
main results of this paper is given by our Eq. \eqref{massless field arbitrary spin u r part in utilde}, i.e.,
\begin{align*}
 \frac{\mathbf{\Psi}^{(\mathbb{s})}_{\ell 0}(u,r)}{\sqrt{2\pi}}=&\delta(\tilde u)
+\left[\frac{\lambda}{2r}+\frac{M}{2r^2}(\mathbb{s}+1)(2\mathbb{s}+1)\right]\theta(\tilde u)
\nonumber\\+&\frac{1}{8r^2}(\ell-\mathbb{s})(\ell-\mathbb{s}-1)(\ell
+\mathbb{s}+1)\nonumber\\\times&(\ell+\mathbb{s}+2)\tilde u\theta(\tilde u)\nonumber\\+&O(r^{-3}).
\end{align*}
It describes the full spectrum of a retarded spin-weighted field in function of 
the Schwarzschild 'corrected' retarded time $\tilde u=u-2M\ln2M$ and the radial 
coordinate $r$, at order $O(1/r^2)$ (sub-leading orders are given in the Appendix) in 
Schwarzschild spacetime. Its structure is then deeply investigated and compared with 
the literature in the case of a scalar field. For perturbations of spin 
$\mathbb{s}\in\{0,\pm2\}$ we obtain a result which is comparable with observations 
coming from LIGO and VIRGO experiments: a spin-independent prompt signal on the null 
cone (given by the $\delta(\tilde u)$ term), consistent with the GW170817 speed bound, 
and tails too small to be observed. We therefore started from a purely mathematical 
observation, which is now a reliable tool able to connect the analytic prescription of 
asymptotic fields to what is often observed in experiments. The paper structure is as follows.

In Sec. ~\ref{Poincare asymptotic expansion} we review the Poincar\'e asymptotic 
expansion technique, deeply investigated in our previous 
papers in Refs \cite{Esposito:2025jsq, Esposito:2026las}. 
In Sec. ~\ref{The Huygens' principle and the decomposition of the Green function} 
we discuss some properties satisfied by the solutions of the wave equation in curved 
spacetime. In particular, we focus on the Huygens' principle and 
analytical properties of the Green function, which are well known in the mathematical 
literature, underlying several important physical features.
In Sec. ~\ref{Scalar perturbations in Schwarzschild spacetime}, in order to provide a 
self-contained presentation, we recall some basics on scalar perturbations in Schwarzschild 
spacetime, a classical topic in black hole physics. We then compute the full spectrum of 
a massless scalar field in Sec. ~\ref{FULL SPECTRUM OF A MASSLESS SCALAR FIELD in schwarzschild}, while in Sec. 
~\ref{The full spectrum of a massive scalar field schwarzschild} we investigate the 
massive case, in order to better understand the role of the mass. 
In Sec. ~\ref{Gravitational waves}, once understood properly the scalar case, we focus 
on massless spin-weighted perturbations. In Schwarzschild spacetime, by virtue of 
spherical symmetry, the spin enters the equations as a mere parameter, modifying only 
the angular part of the field which is however still under control. 
After a brief but detailed comparison with the literature in Sec. 
\ref{sec: a brief comparison with the literature}, we plot in Sec.
~\ref{section plots} all the main solutions we find, offering a direct 
comparison between the scalar case and gravitational waves, enhancing the 'prompt' 
and 'tail' structure discussed in the core of the paper. In Sec. ~\ref{section observations} 
we compare our formulae with the constraints from LIGO and VIRGO observations. In the Appendix we provide all details concerning the 
computations shown in the main part of the paper, more sub-leading orders and a comment 
on the action of superdilations (cf. Ref. \cite{Refuto:2026ljz}) on the asymptotic form of the fields.

\section{The Poincar\'e asymptotic expansion}
\label{Poincare asymptotic expansion}

Confluent Heun differential equations  belong to a large class of ordinary 
differential equations which can be understood as a generalisation of hypergeometric 
differential equations  (cf. Ref. \cite{Ronveaux1995HeunsDE}). Asides the mathematical 
interest, there exist several physical reasons which suggest to investigate these 
equations. Indeed, the dynamics of fields on black hole spacetimes leads to (thanks 
to separation of variables, cf. Sec. 3.2 in Ref. \cite{Esposito:2026las}) 
confluent Heun equations describing the radial part of the perturbation. 
A clear physical interpretation is given, e.g., in Ref. \cite{Minucci:2024qrn}. 
This section aims at giving a mathematical introduction to the problem, highlighting 
the mathematical framework and the reason why we introduce the Poincar\'e asymptotic expansion. 
The notation is detached from the rest of the paper, making it a self-contained interlude.

Confluent Heun differential equations exhibit the general form
\begin{align}
\left[\frac{d^{2}}{dz^{2}}
+\left(\alpha+\frac{(\beta+1)}{z}
+\frac{(\gamma+1)}{(z-1)}\right)\frac{d}{dz} \right. \nonumber\\\left.
+ \left(\frac{\mu}{z}
+\frac{\nu}{(z-1)}\right) \right]Y(z)=0,
\label{confluent Heun equation in its general form}
\end{align}
where the parameters $\mu$ and $\nu$ obey the relations
\begin{align}
\mu=&\frac{1}{2}[\alpha-\beta-\gamma-2 \eta+(\alpha - \gamma)\beta],\\
\nu=&\frac{1}{2}[\alpha+\beta+\gamma+2 \delta + 2 \eta
+(\alpha + \beta) \gamma],
\end{align}
while $\alpha,\beta,\gamma,\delta,\eta$ are free (in general complex) parameters. 
The dynamics of fields in curved backgrounds is often governed by this mathematical 
framework. However, there is one important caveat, regarding the point at infinity, 
i.e., the solution at $z\rightarrow \infty$. In such a neighborhood, Eq. \eqref{confluent 
Heun equation in its general form} is solved by the so-called Thom\'e solution,
\begin{equation}
Y_{\infty}(z)= C_{1}z^{-\theta_{1}} \sum_{k=0}^{\infty} \rho_{k}z^{-k}
+ C_{2}e^{\theta_{2}z} \; z^{-\theta_{3}} \;
\sum_{k=0}^{\infty}\sigma_{k}z^{-k},
\end{equation}
where $C_i$, $\theta_i$ are parameters and $\rho_k$, $\sigma_k$ coefficients that 
must be computed through recursion relations obtained by inserting such a solution 
into Eq. \eqref{confluent Heun equation in its general form}. By virtue of the complexity 
of the task, in general only the leading term is considered. In order to achieve a 
more precise computation of the solution at infinity, we address the problem from a 
different perspective, relying on the work of Poincar\'e on ordinary differential 
equations. The seminal papers are in Refs \cite{Poincare1, Poincare2}, while a more 
complete discussion is given in Sec. II of Ref. \cite{Esposito:2026las}. Here, we 
briefly cover the formalism in a self-contained review. Let us just consider a second-order ordinary 
differential equation with variable coefficients,
\begin{equation}
\left[\frac{d^{2}}{dr^{2}}+p(r)\frac{d}{dr}
+q(r)\right]R(r)=0.
\label{second order radial equation in the right form for poincare}
\end{equation}
It is possible to reduce it to a form where the first derivative is absent, usually known 
as Laguerre-Forsyth  canonical form (cf. Refs \cite{Wilc, Esposito2017}). Upon assuming 
\begin{equation}
    R(r)=\alpha(r)\beta(r),
\end{equation}
the method yields
\begin{equation}
    \alpha(r)={\rm exp} \left(-\frac{1}{2}\int p(r)dr \right),
    \label{canonical alpha piece}
\end{equation}
jointly with the second-order differential equation
\begin{equation}
\left[\frac{d^{2}}{dr^{2}}+J(r)\right]\beta(r)=0,
\label{canonical form of the radial equation}
\end{equation}
where the potential term is given by
\begin{equation}
    J(r)=-\frac{1}{2}\frac{dp}{dr}
-\frac{1}{4}p^{2}(r)+q(r).
\label{canonical potential}
\end{equation}
The advantage of this form is that now we can consider the asymptotic expansion 
of the canonical potential $J(r)$ for large $r$,  
\begin{equation}
    \left[\frac{d^{2}}{dr^{2}}+\sum_{k=0}^{\infty}
\frac{A_{k}(\dots)}{r^{k}}\right]\beta(r)=0,
\label{canonical form of the radial equation at infinity}
\end{equation}
where the $A_k$ coefficients are constant with respect to $r$ but in general might depend 
on several parameters denoted by dots, such as the charge or mass of the field, as well 
as the mass or angular momentum of the black hole.

At~this stage, Eq. \eqref{canonical form of the radial equation} 
suggests looking for $\beta(r)$ at large $r$ in the form (setting 
$\varepsilon= \pm 1$ and considering 
$\zeta$ as a parameter to be determined as shown~below)
\begin{equation}
\beta(r) \sim e^{\varepsilon r \sqrt{-A_{0}}} \; r^\zeta
\; \left(1+\sum_{s=1}^{\infty}
\frac{B_{s}(\dots)}{r^{s}}\right).
\label{canonical beta function poincare}
\end{equation}
where $\varepsilon=\pm 1$ and $\zeta$, $B_s(\dots)$ are coefficients that can be 
obtained by inserting this solution into Eq. \eqref{canonical form of the radial equation at infinity}.
Moreover, since we look for a bounded solution at large $r$, we restrict ourselves 
to the $\varepsilon=-1$ case (the $\varepsilon=1$ mode cannot be avoided in studying, e.g., 
the behavior near the horizon). After a straightforward computation we obtain 
(in the $\varepsilon=1$ case, $\zeta$ flips sign):
\begin{align}
    \zeta=&\frac{A_1}{2\sqrt{-A_0}}, 
    \label{zeta general expression}\\
    B_1=&\frac{\zeta(\zeta-1)+A_2}{2(\zeta-1)\sqrt{-A_0}-A_1}, 
    \label{B1 general expression}
\end{align}
while the general recursion formula is given by
\begin{align}
    2\sqrt{-A_0}sB_s=-&\Big[(\zeta-s+1)(\zeta-s)B_{s-1}\nonumber\\&\sum_{j=2}^{s+1}A_jB_{s+1-j}\Big]
\end{align}

We do not write explicitly further coefficients because they have a cumbersome form. However, 
they can be easily computed with symbolic algebraic programs, such as Mathematica or Maple. 
Note that the $B_s$ coefficients depend on the same parameters of $A_k$, obviously in a 
different manner. We report for completeness the full radial solution, given by 
\begin{equation}
    R(r)=\mathcal{N}\alpha(r)e^{- r \sqrt{-A_{0}}} \; r^\zeta
\; \left(1+\sum_{s=1}^{\infty}
\frac{B_{s}(\dots)}{r^{s}}\right),
\label{full radial asymptotic solution}
\end{equation}
where $\mathcal{N}$ is a normalization constant. In light of the above, the algorithm 
provided is well-suited for computer programs and makes it possible to obtain a solution with the 
desired level of accuracy, going beyond the leading order (which is usually similar to 
the case $B_s=0 $ for every $s$) so often considered in the literature.

\subsubsection{Addendum: the relation with Whittaker functions}

In Subsec. ~\ref{sec: a brief comparison with the literature} we compare Eq. 
\eqref{full radial asymptotic solution}, computed for a spin-weighted field, 
with typical beyond-the-leading-order analyses exploited in the literature, 
relying on Whittaker functions. The connection between our formalism and such a 
class of functions was first noted by the authors in Ref. \cite{Esposito:2025jsq}. 
In this interludio we aim at formalising it in order to give a dictionary among 
these two methods, highlighting differences. The main reference, besides standard 
textbook on higher transcendental functions (cf. Ref. \cite{Erdelyi:1953hft}), 
is given by Chap. XIII of Ref. \cite{NIST:DLMF}. The Whittaker's equation is 
nothing but Eq. \eqref{canonical form of the radial equation at infinity} truncated 
at $k\le2$, which is written in the mathematical literature as ($z$ being a complex variable)
\begin{equation}
\left[\frac{d^2}{dz^2}-\frac{1}{4}+\frac{k_W}{z}+\frac{\frac{1}{4}-\mu_W^2}{z^2}\right]W(z)=0.
\end{equation}
We can obtain this standard form from Eq. \eqref{canonical form of the radial equation at infinity} 
upon rescaling $z=2\sqrt{-A_0}r$ and considering as Whittaker indices $k_W=\zeta$ and $1/4 -\mu_W^2=A_2$, 
showing a regular singular point at the origin with indices $\frac{1}{2}\pm \mu_W$ and an 
irregular one at infinity, of rank $1$. Its solutions, $M_{k_W,\mu_W}(z)$ and $W_{k_W,\mu_W}(z)$, 
are given in terms of Kummer confluent hypergeometric functions, whose explicit form 
is not needed. The solution which is exponentially non-growing at infinity is $W_{k_W,\mu_W}(z)$, 
whose asymptotic form is
\begin{align}
W_{k_W,\mu_W}(z)\sim e^{-\frac{z}{2}}z^{k_W}\sum_{s=0}^\infty &\frac{1}{s!}\left(
\frac{1}{2}+\mu_W-k_W\right)_s\nonumber\\& \times\left(\frac{1}{2}-\mu_W-k_W\right)_s(-z)^{-s},
\end{align}
where the $(a)_s=\Gamma(a+s)/\Gamma(a)$ are the Pochhammer's symbol. Note that such 
a function resembles the $\beta(r)$ solution in Eq. \eqref{canonical beta function poincare}, 
once the outgoing wave condition (i.e., $\varepsilon=-1$) is imposed. We have therefore 
that the standard mathematical theory gives a reliable framework in computing asymptotic 
fields, provided that one stops at the $1/z^2$ term in the starting equation. However, as 
we point out in the following, several physical features are encoded in the sub-leading terms, 
which change qualitatively the aspect of the field. 

\section{Mathematical properties of the wave equation}
\label{The Huygens' principle and the decomposition of the Green function}

The study of the propagation of fields in curved background has its roots in a plethora 
of mathematical and numerical results, collected and enhanced over the last decades. 
In accordance with our purposes, we give a brief description of some of them.

\subsection{The Huygens' principle}

With the metric signature $(-,+,+,+)$ (the same of Ref. \cite{Esposito:2025jsq}), 
the wave equation on a curved spacetime reads as
\begin{equation}
     (\Box-\mu^2)\Phi(x)=0.
\end{equation}
We first recall a mathematical property of such a differential wave operator
and the related Cauchy problem. A second order linear partial differential equation of normal 
hyperbolic type is said to satisfy Huygens’ principle if, for any Cauchy problem, the 
value of the solution at any spacetime point $P$ depends only on the data on the 
intersection $C^\pm(P)\cap M$ of the future (past) characteristic conoid with the 
initial manifold $M$. The condition on both $C^\pm(P)$ relies, in physical terms, 
on the time-reversal symmetry of the differential equation. The support of these 
fundamental solutions of, in this case, the wave equation, is therefore restricted 
to both $C^+(P)$ and $C^-(P)$, without terms propagating inside or outside the 
characteristic conoid, i.e., without \textit{tails} (cf. the seminal papers in Refs 
\\ \cite{AIHPA_1974__20_2_153_0, AIHPA_1974__21_4_347_0} or Sec. 5.7 in 
Ref. \cite{Friedlander:2010eqa}, for some remarks on inequivalent formulations of the Huygens' 
principle, cf. Ref. \cite{Bombelli:1994rh}). The identification of differential equations 
satisfying such a principle is a rich mathematical research line, having produced 
several results of physical interest. Of particular interest for our aims is Theorem 
5.7.2 in Ref. \cite{Friedlander:2010eqa}, which states that the wave equation 
$\Box\psi=0$ in empty spacetime satisfies Huygens' principle (cf. Ref 
\cite{McLenaghan_1969}) if and only if the spacetime is flat (trivial case) or a 
plane wave spacetime where other conditions must hold. Since we are considering a 
Petrov D spacetime, we expect from a full spectrum analysis over the field solution of 
Eq. \eqref{wave equation massive} in the massless case, that the solution propagates 
not only on the lightcone, but also\footnote{The propagation of the field 
solution outside the lightcone would violate causality.} inside it. 

\subsection{The Green function of the wave equation}

From a physical perspective, 
a sharp signal fails to propagate sharply. Wave fronts do not evolve 
at the speed of light, but they cover all possible velocities (tail terms inside the 
lightcone) due to the interaction with the spacetime curvature. In the physical literature, 
such a signature is widely recognized through the study of Green functions. 
The discussion in Refs  \cite{DeWitt:1960fc, HOBBS1968141} of the tail 
term in the curved spacetime retarded Green function of a moving charged mass point (cf. 
Eq. $(2.16)$ in Ref. \cite{DeWitt:1960fc}) is remarkable: indeed, the electromagnetic Green function 
exhibits a tail on any initially sharp pulse of radiation. Such a result extends to scalar 
and gravitational perturbations too in a similar fashion. In order to provide a brief, 
self-contained  although precise introduction of the topic aligned with our purposes, 
we follow the thorough review in Ref. \cite{Poisson:2011nh}. 

\subsubsection{Green function for a massive scalar field in flat spacetime}

A massive scalar field $\Phi(x)$ in flat spacetime satisfies the inhomogeneous wave equation
\begin{equation}
    (\Box-\mu^2)\Phi(x)=-4\pi\rho(x),
\end{equation}
where the Green function $G(x,x')$, defined as 
\begin{equation}
    \Phi(x)=\int d^4x' \rho(x')G(x,x'),
\end{equation}
satisfies the differential equation
\begin{equation}
    (\Box-\mu^2)G(x,x')=-4\pi \delta_4(x-x').
\end{equation}
Of particular interest are the retarded and advanced Green functions, with the 
property that they vanish when $x$ is in the past (resp. the future) of $x'$. Upon 
considering Lorentz invariance and homogeneity of spacetime, they can be written as
\begin{equation}
    G_\text{ret}(x,x')=\theta(t-t')g(\sigma), \quad  G_\text{adv}(x,x')=\theta(t'-t)g(\sigma),
\end{equation}
where
\begin{equation}
    \sigma=\frac{1}{2}\eta_{ab}(x-x')^a(x-x')^b
    \label{synge world function in flat spacetime}
\end{equation}
is Synge's world function in flat spacetime and $g(\sigma)$ a function to be 
determined. Through distributional analysis techniques, we find that
\begin{equation}
g(\sigma)=\delta(\sigma)-\frac{\mu}{\sqrt{-2\sigma}}J_1\left(\mu\sqrt{-2\sigma}\right)\theta(-\sigma),
\label{scalar green function massive field flat spacetime}
\end{equation}
where $J_1(z)$ is a Bessel coefficient (cf. Sec.  ~\ref{appendix section bessel functions}).

This and all the subsequent results cited or derived in this paper must be understood 
as distributions. In this context, the convention on the value of $\theta(0)$ does 
not really matter (as understood from Ref. \cite{Poisson:2011nh}), since it leads 
to ambiguity only on a zero-measure set of points. Moreover, it is understood as a 
limiting procedure such as $\lim_{\varepsilon\rightarrow 0}\theta(\varepsilon)$, avoiding 
coincidences with a specific prescription. 

\subsubsection{Green functions in curved spacetime and their decomposition}

In a curved spacetime, the dynamics of a massless scalar field is governed by the wave equation
\begin{equation}
    (\Box-\xi R)\phi(x)=-4\pi \rho(x),
\end{equation}
where $\xi$ is the conformal coupling to the Ricci scalar $R$. The Green function $G(x,x')$ such that
\begin{equation}
    \phi(x)=\int d^4x'\sqrt{-g'}G(x,x')\rho(x'),
\end{equation}
satisfies the point-source wave equation
\begin{equation}
    (\Box-\xi R)G(x,x')=-4\pi \delta_4(x,x'),
\end{equation}
where the invariant Dirac $\delta$ is defined as $\delta_4(x,x')=(gg')^{-1/4}\delta_4(x-x')$, 
where $\delta_4(x-x')=\Pi_{i=0}^3 \delta(x^i-x'^i)$. After denoting by $\sigma(x,x')$ the
Synge world function, defined as half the square of the geodesic distance between $x$
and $x'$, we need to define two more useful quantities.
The light cone step function $\theta_\pm(-\sigma)$, where $\sigma(x,x')$ 
is the curved spacetime world function, is defined as
\begin{equation}
    \theta_\pm(-\sigma)=\theta_\pm(x,\Sigma)\theta(-\sigma), \quad x'\in\Sigma,
\end{equation}
where $\theta_+(x,\Sigma)=1$ when $x$ belongs to the future of the spacelike hypersurface 
$\Sigma$ and zero otherwise, while $\theta_-(x,\Sigma)=1-\theta_+(x,\Sigma)$. Briefly, we have
\begin{equation}
    \theta_\pm(-\sigma)=\begin{cases}
        1 \quad x\in I^\pm(x'),\\ 0 \quad \text{otherwise}.
    \end{cases}
\end{equation}
Moreover, $\theta_+(-\sigma)+\theta_-(-\sigma)=\theta(-\sigma)$. Thanks to such a 
distribution, we can define the lightcone Dirac functional as
\begin{equation}
    \delta\pm(\sigma)=\theta_\pm(x,\Sigma)\delta(\sigma), \quad x'\in\Sigma.
\end{equation}
We have therefore that $\delta_\pm(\sigma)$ is supported, as function of $x$, on the 
future (past) lightcone of $x'$. The property $\delta_+(\sigma)+\delta_-(\sigma)=\delta(\sigma)$ holds. 
The retarded and advanced solutions $G_\pm(x,x')$ (as a function of $x$, nonvanishing in the 
causal future, resp. past, of $x'$), throughout the Hadamard construction 
(cf. Ref. \cite{jacques2020lectures}) yields 
\begin{equation}
    G_\pm(x,x')=U(x,x')\delta_\pm(\sigma)+V(x,x')\theta_\pm(-\sigma).
\end{equation}
where $U(x,x'), V(x,x')$ are biscalars, whose definition is given in Ref. 
\cite{Poisson:2011nh} (cf Subsec. 14.2). However, these are not the main important 
quantities. As outlined in Ref. \cite{PhysRevD.67.024025}, the curved spacetime 
Green function can be decomposed into two pieces. The singular one, solution of the equation
\begin{equation}
    (\Box-\xi R)G_S(x,x')=-4\pi\delta_4(x,x'),
\end{equation}
is given by
\begin{equation}
    G_S(x,x')=\frac{1}{2}U(x,x')\delta(\sigma)-\frac{1}{2}V(x,x')\theta(\sigma).
    \label{singular green function}
\end{equation}
It does not distinguish between past and future and its support excludes $I^\pm(x')$, where $\theta(\sigma)=0$.
The regular piece satisfies
\begin{equation}
    (\Box-\xi R)G_R(x,x')=0,
\end{equation}
and reads as
\begin{align}
    G_R(x,x')=&G_+-G_S\\=&\frac{1}{2}U(x,x')\left[\delta_
+(\sigma)-\delta_-(\sigma)\right]\nonumber\\&+V(x,x')
\left[\theta_+(-\sigma)+\frac{1}{2}\theta(\sigma)\right].
    \label{regular green function}
\end{align}
This reveals directly that
\begin{align}
    G_R(x,x')=G_+(x,x') \quad x\in I^+(x'),
\end{align}
since, in that case, $\theta(\sigma)=0$ and $\theta_+(-\sigma)=1$. Furthermore, 
its support does not include $I^-(x')$, where $\theta(\sigma)=\theta_+(-\sigma)=0$.
As a last remark, the Hadamard construction of the Green functions is valid for 
$x$ belonging to a normal convex neighbourhood of $x'$, where the Synge world 
function $\sigma$ is single-valued (cf. Sec. XIV in Ref. \cite{Poisson:2011nh}).

\subsubsection{Green functions in black hole theory}

The mathematical machinery briefly introduced helps us in the study of wave propagation 
in black hole spacetimes, where the regular and singular part of the Green 
function acquires a clear physical meaning.

In Ref. \cite{Leaver:1986gd}, Leaver analysed the massless Green function for 
arbitrary spin fields (labelled with $s$) in Schwarzschild spacetime. Such a 
function solves the equation (resembling Eq. \eqref{radial equation schwarzschild regge wheeler})
\begin{align}
&\left[\frac{\partial^2}{\partial r_*^2}-\frac{\partial^2}
{\partial t^2}\right. \nonumber\\&\left. -f(r)\left(\frac{\ell(\ell+1)}{r^2}
+(1-s^2)\frac{2M}{r^3}\right)\right]G(x,x')=\delta_2(x,x'),
\label{massless green function schwarzschild rstar t arbitrary spin}
\end{align}
where in this case $x=\{t,r_*\}$ and $r=r(r_*)$. It turns out that $G(x,x')$ can be 
decomposed into a sum over quasinormal modes, an integral around the branch cut in 
the frequency parameter, and an integral over asymptotically large magnitudes of the 
frequency, resembling the sharp pulse and tail structure. A clear mathematical 
description of these pieces can be given by analysing the Fourier transform of the Green 
function in the complex plane. Quasinormal modes are solutions concerning the poles of 
the Green function, while the integral around the branch cut gives rise to the tail 
contribution, strongly dependent on the form of the potential. The potential in Eq. 
\eqref{massless green function schwarzschild rstar t arbitrary spin} leads to the 
well-known Price power-law decay, $t^{-(2\ell +3)}$ (cf. Refs \cite{Price:1971fb, PhysRevD.5.2439}), 
which is remarkably independent of the spin of the field. The prompt contribution 
is instead given by the large-$\omega$ behavior of the Fourier transform of $G(x,x')$. 
Such properties are well described in Ref. \cite{PhysRevD.52.2118}, where numerical 
computations support the analytical treatment of the problem. From a broader perspective, 
the analytical and numerical analysis in Ref. \cite{Zenginoglu:2012xe} captures several 
features of scalar wave propagation in Schwarzschild spacetime. Furthermore, of 
central importance is the geometrical optics approximation (valid for $\omega\rightarrow \infty$) 
of the Fourier transform of the Green function,
\begin{align}
G(\omega;\mathbf{x},\mathbf{x}')\simeq \sum_{n=0}^\infty \frac{A_n(\mathbf{x},\mathbf{x}')}
{(i\omega)^n}e^{i\omega T(\mathbf{x},\mathbf{x}')},
\end{align}
where $T(\mathbf{x},\mathbf{x}')$ is called the eikonal (wavefronts are its level 
surfaces). Such a structure resembles ours.

\section{The wave equation in Schwarzschild spacetime}
\label{Scalar perturbations in Schwarzschild spacetime}

The dynamics of fields in black hole spacetimes is described by partial 
differential equations in curved Lorentzian manifolds. In this mathematical 
framework the Poincar\'e asymptotic series introduced in Sec. ~\ref{Poincare asymptotic expansion} 
aims at describing such fields in the asymptotic region of these spacetimes. 
We devoted our first papers (cf. Refs \cite{Esposito:2025jsq}, \cite{Esposito:2026las}) 
to this kind of investigations, starting with Schwarzschild spacetime and then 
extending our results to every Petrov-D metric. Since we are now interested in a 
deeper analysis in Schwarzschild spacetime, we recall some basics in order to set the notation and conventions.
An overview on black hole perturbation theory and its physical applications can be 
found in Ref. \cite{Cardoso:2025npr} (and references therein). 
In our investigations, we consider the spacetime as a fixed background, neglecting the fields back-reaction.

In~standard spherical coordinates $(t,r,x^A)$ where 
$t \in {\mathbb{R}} \cup \{-\infty \} \cup \{+\infty \}, r \geq 0,
x^A=(\theta,\phi) \in [0,\pi]\times [0,2\pi)$, with $G=c=1$ units, upon defining
\begin{equation}
    f(r)=1-\frac{2M}{r}, 
    \label{1-2M on R schwarzschil factor}
\end{equation}
the~spacetime metric of a Schwarzschild black hole reads as
\begin{equation}
g_{S}=-f(r) dt^{2}+\frac{dr^{2}}{f(r)}+r^{2}(d\theta^{2}+\sin^{2}\theta d\phi^{2}).
\end{equation}
The wave equation
\begin{equation}
    (\Box-\mu^2)\psi_\mu=0
    \label{wave equation massive}
\end{equation}
for a massive scalar field $\psi_\mu(t,r,x^A)$ can be solved by noticing that 
the wave operator can be written as (here $g$ is the determinant of the matrix
of metric components) 
\begin{equation}
    \Box=\frac{1}{\sqrt{-g}}\partial_a(\sqrt{-g}g^{ab}\partial_b)
\end{equation}
and by considering spacetime symmetries. Upon exploiting spherical symmetry, we can expand 
the $\psi_\mu$ over monochromatic plane waves as
\begin{equation}
    \psi_\mu=e^{-i\omega t}\sum_{\ell=0}^\infty \frac{\mathcal{R}_{\ell\omega \mu}(r)}{r}
\sum_{m=-\ell}^\ell Y^m_\ell (x^A).
\label{standard separation of variables in the litearute}
\end{equation}
Furthermore, the wave equation can be simplified by considering the tortoise 
coordinate $r_*$, such that $dr_*/dr=1/f(r)$, 
\begin{equation}
r_*=r+2M\ln\left(\frac{r}{2M}-1\right).
\label{Schwarzschild tortoise coordinate}
\end{equation}
Eq. \eqref{wave equation massive} therefore reduces to the 
one-dimensional radial differential equation
\begin{equation}
    \left[\frac{d^2}{dr^2_*}+\omega^2-V(r_*)\right]\mathcal{R}_{\ell\omega \mu}(r_*)=0
    \label{radial equation schwarzschild regge wheeler}
\end{equation}
with effective potential
\begin{equation}
    V(r_*)=f(r)\left[\frac{\ell(\ell+1)}{r^2}+\frac{2M}{r^3}+\mu^2\right],
    \label{massive regge wheeler potential}
\end{equation}
where $r$ must be understood as $r(r_*)$.

This framework is the core of black hole perturbation theory (cf. Ref. 
\cite{Chandrasekhar:1985kt}), holding for arbitrary-spin fields with minor 
changes. The major part of the literature concerns such a decomposition, which can 
be performed in every Petrov D metric (cf. Ref. \cite{Esposito:2026las}).

\subsection{The Poincar\'e asymptotic expansion}

In our framework we recover the a priori $1/r$ factor of the field expansion in Eq. 
\eqref{standard separation of variables in the litearute}, but only after having 
considered the decomposition coming from the Laguerre-Forsyth canonical reduction. 
A generic field (we highlight the mass in our notation) can be written as
\begin{equation}
\psi(t,r,x^A;\mu)=\sum_{\ell=0}^{\infty}\psi_{\ell \mu}(t,r)\sum_{m=-\ell}^{\ell}
C_{\ell m}Y_{\ell m}(x^A), 
\label{our generic field decomposition}
\end{equation}
where $C_{\ell m}$  are the expansion coefficients of spherical harmonics and 
$\psi_{\ell \mu}$ are the integrated Fourier modes
\begin{equation}
\psi_{\ell \mu}(t,r)=\int_{-\infty}^{\infty}\frac{d\omega}{\sqrt{2\pi}}e^{-i \omega t}
R_{\ell \omega \mu}(r).
\end{equation}
The purely radial function $R_{\ell\omega \mu}$ solves Eq. \eqref{second order 
radial equation in the right form for poincare} with
\begin{align}
p(r)=&\frac{2(r-M)}{r^{2}f(r)},\\
q_{\ell \omega \mu}(r)=&\frac{1}{f(r)}\left(\frac{\omega^{2}}{f(r)}-\frac{\ell(\ell+1)}{r^{2}}
-\mu^{2}\right).
\end{align}

In Schwarzschild geometry, Eq. \eqref{canonical alpha piece} gives
\begin{align}
\alpha(r)=&\frac{1}{\sqrt{r(r-2M)}}\nonumber\\=&r^{-1}f(r)^{-1/2},
\end{align}
while the canonical potential in Eq. \eqref{canonical potential} reads as 
\begin{align}
    J_{l\omega \mu}(r)=& \frac{1}{4r^{2}}+\frac{1}{4(r-2M)^{2}}
-\frac{\left[\frac{1}{2}+\ell(\ell+1)\right]}{r(r-2M)}
\nonumber \\
+&\frac{\omega^{2}r^{2}}{(r-2M)^{2}}
-\frac{\mu^{2}r}{(r-2M)}.
\end{align}
A detailed analysis of this potential and related local solutions near 
the singular points $r=0,2M$ can be found in Ref. \cite{Esposito:2025jsq}.
For this potential, the coefficients of the asymptotic expansion (as written 
in Eq. \eqref{canonical form of the radial equation at infinity}) are 
\begin{align}
   A_0=&\omega^2-\mu^2,\\
  A_k=&(2M)^k[(k+1)\omega^2-\mu^2]\nonumber\\+&\left[\frac{1}{4}\delta_{k2}
+\left(\frac{k-3}{4}-\ell(\ell+1)\right)(2M)^{k-2}\right]_{k\ge2} \quad k\ge1.
\end{align}
Upon applying the mathematical machinery described in Sec. ~\ref{Poincare asymptotic expansion}, 
the solution of Eq. \eqref{wave equation massive} resembles the structure of Eq. \eqref{standard 
separation of variables in the litearute} (upon explicitly factorising the $1/r$ term)  and  for $\mu=0$ is given by
\begin{equation}
\psi(t,r,x^A;0)=\mathcal{N}\sum_{\ell=0}^\infty\frac{     \mathbf{\Psi}_{\ell 0}(t,r)}{r}
\sum_{m=-\ell}^\ell C_\ell^m Y_\ell^m(x^A),
\label{Schwarzschild full massless field}
\end{equation}
where
\begin{equation}
     \mathbf{\Psi}_{\ell 0}(t,r)=f(r)^{-\frac{1}{2}}\psi_{\ell 0}(t,r)
     \label{mathbf psi scalar}
\end{equation}

and the integrated Fourier modes read as
\begin{align}
\psi_{\ell0}(t,r)=\int_{-\infty}^\infty \frac{d\omega}{\sqrt{2\pi}}&
e^{-i\omega t-r\sqrt{-\omega^2}+\frac{2\omega^2 M}{\sqrt{-\omega^2}}\ln r}
\nonumber\\\times&\left(1+\sum_{s=1}^{\infty}\frac{B_s(\ell,\omega;0)}{r^s}\right).
\label{integrated fourier modes}
\end{align}
Such an integral is the core of our investigations.

\section{Properties of the full spectrum of a massless scalar field}
\label{FULL SPECTRUM OF A MASSLESS SCALAR FIELD in schwarzschild}

We aim at computing explicitly the integral in Eq. \eqref{integrated fourier modes}, 
in order to obtain an analytic form of the full spectrum of the field. Moreover, 
for phenomenological (and mathematical, since the series is an asymptotic expansion) 
reasons we are interested in the asymptotic region, in particular future null 
infinity $\mathcal{I^+}$, aiming at describing an outgoing massless scalar wave. 
Therefore, we choose the branch cut\footnote{Upon 
choosing the branch cut $\sqrt{-\omega^2}=i\omega$, the analysis at past null 
infinity $\mathcal{I}^-$ follows straightforwardly due to the introduction of 
the advanced time $v=t+r_*$.}
\begin{equation}
\sqrt{-\omega^2}=-i\omega
\end{equation}
and we introduce Bondi coordinates $(u,r,x^A)$, where 
$u\in(-\infty,\infty)$ is the retarded time defined as
\begin{equation}
u=t-r_*.
\label{retarded time Schwarzschild}
\end{equation}
In the Schwarzschild case, such a coordinate system coincides with retarded 
Eddington-Finkelstein coordinates and is well-suited in the study of outgoing 
waves (ingoing waves are easily described through the advanced time $v=t+r_*$). 
The field can then be rewritten as\footnote{The principle of general covariance 
ensures that such a change in coordinates can be performed directly on the solutions 
of the equation of motions, without losing any information.}
\begin{align}
    \psi_{\ell 0}(u,r)=\int_{-\infty}^\infty &\frac{d\omega}{\sqrt{2\pi}}
e^{-i\omega \left(u-2M\ln \frac{2Mr}{r-2M}\right)}\nonumber\\&\times 
\left(1+\sum_{s=1}^{\infty}\frac{B_s(\ell,\omega;0)}
{r^s}\right).
\label{Schwarzschild u-r massless field}
\end{align}
One important feature of the massless case is that the integral over the whole 
spectrum can be computed exactly. Indeed, since the $B_s(\ell,\omega;0)$ coefficients 
are Laurent polynomials\footnote{An interesting remark is that, for $M=0$, the 
$B_s(\ell,\omega;0)$ coefficients coincide with the spherical-Hankel ones 
$i^k(\ell+k)!/[k!(\ell-k)!(2\omega)^k]$ and the series terminates at 
$k=\ell$.} in $\omega$, the integral structure underlying our analysis is provided by
\begin{equation}
 I_n=\int_{-\infty}^\infty \frac{d\omega}{\sqrt{2\pi}}e^{-i\omega 
\left(u-2M\ln \frac{2Mr}{(r-2M)}\right)} \omega^n, \quad n\in \mathbb
    Z.
    \label{the massless In integral}
\end{equation}
Therefore, as explained in the Appendix, we derive the exact result ($\delta^{(n)}$ 
labels the $n$-th derivative of the Dirac $\delta$, $\delta^{(0)}=\delta$)
\begin{equation}
    I_n=
    \begin{cases}
        i^n \sqrt{2\pi} \delta^{(n)}[A] \quad n\ge0;\\
        i^n\sqrt{2\pi}\frac{\theta(A)}{(|n|-1)!}A^{|n|-1}, \quad n<0,
        \label{evaluation of the massless In integral}
    \end{cases}
\end{equation}
where 
\begin{equation}
   A= A(u,r)=u-2M\ln \frac{2Mr}{(r-2M)}.
\end{equation}

In order to analyse the full structure of the field in Eq. \eqref{Schwarzschild full massless field}, 
we can consider, e.g., the $s=1,2$ terms in the expansion in Eq. 
\eqref{Schwarzschild u-r massless field}. The coefficients are 
\begin{align}
    B_1(\ell,\omega;0)&=-4iM^2 \omega -M+\frac{i}{2\omega}\ell(\ell+1),\\
    B_2(\ell,\omega;0)&=-8\omega^2M^4+2M^2\left[\ell(\ell+1)-\frac{1}{4}\right]
\nonumber \\&-\frac{iM}{2\omega}[\ell(\ell+1)-1]-\frac{\ell}{8\omega^2}(\ell^2-1)(\ell+2).
\end{align}
As expected, the $B_s$ coefficients, with respect to $\omega$, exhibit a regular 
and a singular part, following the structure outlined in Sec. ~\ref{The Huygens' 
principle and the decomposition of the Green function} for the Green function 
(cf. Eqs \eqref{regular green function}, \eqref{singular green function}). 
Equation \eqref{Schwarzschild u-r massless field} is then written in terms of the 
generating integral $I_n$ as
\begin{align}
      \psi_{\ell 0}(u,r)=&\left[1-\frac{M}{r}+2\frac{M^2}{r^2}\left[\ell(\ell+1)
-\frac{1}{4}\right]\right]I_0\nonumber\\-&\frac{4iM^2}{r}I_1-\frac{8M^4}{r^2}I_2
\nonumber\\+&\frac{i}{2}\left[\frac{\ell(\ell+1)}{r}-M\frac{\ell(\ell+1)-1}{r^2}\right]I_{-1}
\nonumber\\-&\frac{\ell(\ell^2-1)(\ell+2)}{8r^2}I_{-2}+O(r^{-3}).
\end{align}
Upon considering Eq. \eqref{evaluation of the massless In integral},
we arrive at the final result 
\begin{align}
  \frac{\psi_{\ell 0}(u,r)}{\sqrt{2\pi}}=&\varrho_0(r)\delta(A)+\varrho_1(r)\delta^{(1)}(A)
+\varrho_2(r)\delta^{(2)}(A)\nonumber\\&+2\varsigma_1(r)\theta(A)+2\varsigma_2(r)A\theta(A) +O(r^{-3}),
    \label{Schwarzschild u-r massless field explicit up to r-2}
\end{align}
where the $\varrho_i(r)$ coefficients emerge from the regular part of the $B_s$, 
while $\varsigma_i(r)$ result from the singular one, and we find
\begin{align}
    \varrho_0(r)=&1-\frac{M}{r}+\frac{2M^2}{r^2}\left[\ell(\ell+1)-\frac{1}{4}\right]+O(r^{-3}),\\
     \varrho_1(r)=&\frac{4M^2}{r},\\
      \varrho_2(r)=&\frac{8M^4}{r^2},\\
      \varsigma_1(r)=&\frac14\left[\frac{\ell(\ell+1)}{r}-M\frac{\ell(\ell+1)-1}{r^2}\right],\\
      \varsigma_2(r)=&\frac{\ell(\ell-1)(\ell+1)(\ell+2)}{16\,r^2}.
\end{align}

Two general remarks on Eq. \eqref{Schwarzschild u-r massless field explicit up to r-2}. All 
$A$-dependent pieces are formal: they become smooth, finite quantities once 
smeared against a physical source spectrum, exactly as in Green function language.
Furthermore, since $A(u,r)$ is a real function, $\psi_{\ell 0}(u,r)$ is manifestly 
real: every factor of $i$ carried by the coefficients in the original mode sum is 
exactly cancelled by the $i^n$ built into $I_n$.

The general mathematical features of Sec. ~\ref{The Huygens' principle and the 
decomposition of the Green function} can be observed directly from the 
structure unravelled by such coefficients.
However, before looking directly at the field in Eq. \eqref{Schwarzschild u-r 
massless field explicit up to r-2}, we can collect several informations from 
the two parts of $I_n$ in Eq. \ref{evaluation of the massless In integral}, 
directly connected with the regular and singular part in $\omega$ of the $B_s$ coefficients. 
Indeed, the split between the local $n\ge0$ and global $n<0$ contributions identifies the decomposition of 
the signal between a direct part supported on the null hypersurface $\tilde u=0$  (as clearly depicted by Eq. \eqref{scalar massless field in the corrected retarded time}) and the tail, 
respectively. We can therefore observe the failure of the Huygens' principle 
(described in Sec. ~\ref{The Huygens' principle and the decomposition of the 
Green function}) directly from this computation.
The contributions to the full spectrum coming from the poles of the $B_s$ 
coefficients 
have two origins. The $M$-independent ones, e.g., the $i\ell(\ell+1)/(2\omega)$ term in $B_1$, are present in flat space and give a multipole structure (cf. Eq. \eqref{flat space limit scalar}). The $M$-dependent ones give instead a curvature-induced tail (cf., e.g., the $M\theta(\tilde u)$ term in Eq. \eqref{scalar massless field in the corrected retarded time}), which signals the failure of Huygens' principle in Schwarzschild spacetime,
by virtue of the nonvanishing Weyl curvature part in the Riemann tensor.

In order to further unravel physics features behind Eq. \eqref{Schwarzschild u-r massless 
field explicit up to r-2}, we expand in power series of $1/r$ the $A(u,r)$ function. 
First, we notice that 
\begin{equation}
A=\tilde u+2M\ln f(r),
\end{equation}
where the corrected retarded time (which plays the role of time in our main formulae, 
even if one could define the retarded time in order to have just $\tilde u=u$) is given by
\begin{equation}
\tilde u=u-2M\ln(2M). 
\label{definition of the corrected retarded time}
\end{equation}
We have therefore that 
\begin{equation}
A=\tilde u-\varepsilon(r),
\end{equation}
with 
\begin{equation}
    \varepsilon(r)=\frac{4M^2}{r}+\frac{4M^3}{r^2} +O(r^{-3}).
\end{equation}
In this way we can regard the $\delta^{(n)}(A)$ part of the field in Eq. \eqref{Schwarzschild 
u-r massless field explicit up to r-2} as an expansion leading to $\delta(\tilde u)$, since
\begin{equation}
   \delta(A)+\varepsilon\delta'(A)+\frac{1}{2}\varepsilon^2\delta^{''}(A)+\dots= \delta(\tilde u).
    \label{massless delta utilde in function of delta A}
\end{equation}
This observation allows us to explicitly show the $1/r$ contributions encoded in $A$:
upon considering Eq. \eqref{massless delta utilde in function of delta A}, 
the radial field in Eq. \eqref{Schwarzschild full massless field} (excluding the $1/r$ 
factor as commonly done in the literature) reads as
\begin{align}
\frac{     \mathbf{\Psi}_{\ell 0}(u,r)}{\sqrt{2\pi}}=&\delta(\tilde u)+\frac{1}{2}\left(\frac{\ell(\ell+1)}{r}
+\frac{M}{r^2}\right)\theta(\tilde u)\nonumber\\+&\frac{\ell(\ell+1)
[\ell(\ell+1)-2]}{8r^2}\tilde u\theta(\tilde u)+O(r^{-3}).
\label{scalar massless field in the corrected retarded time}
\end{align}
Such a formula, obtained for a massless field and generalised to spin-weighted fields 
in Sec. ~\ref{Gravitational waves}, represents the core of our present investigation. In 
addition to the clear prompt and tail structure exhibited, it is useful in order to provide 
contacts with phenomenological observations and to study interesting limits.

\subsubsection{Fixed $r$, late-time behavior}

If, instead of null infinity, we want to study the late-time behavior of the 
field, we can consider the $u\rightarrow \infty$ limit in Eq. 
\eqref{Schwarzschild u-r massless field} at fixed $r$, obtaining that only the 
singular terms contribute, since the regular ones are supported only on the 
lightcone. Upon noting that, in such a limit, $A\rightarrow u + cte$, we obtain
\begin{align}
    \psi_{\ell 0}(u,r)\rightarrow\begin{cases}
        0 \quad u<0,\\ 2\sqrt{2\pi}[\varsigma_1(r)+\varsigma_2(r)u] \quad u\ge0.
    \end{cases}
\end{align}
The term linear in $u$ vanishes for $\ell=0,1$, which are exactly the nonradiative 
monopole and dipole multipoles. We are unable to observe the genuine Price power-law 
(cf. Refs \cite{Price:1971fb, PhysRevD.5.2439})
tail by virtue of the truncation of the series. It would be necessary to sum all 
contributions, a task which goes beyond the nature of the method itself.

\subsubsection{Flat-space limit} 

In the flat spacetime limit, i.e., $M\rightarrow 0$, Eq. \eqref{Schwarzschild u-r 
massless field explicit up to r-2} analytically reduces to 
\begin{align}
\frac{\psi_{\ell 0}(u,r)}{\sqrt{2\pi}}\Big|_{M=0}=&
\delta(u)+\frac{\ell(\ell+1)}{2r}\,\theta(u)\nonumber\\&
+\frac{\ell(\ell-1)(\ell+1)(\ell+2)}{8r^2}\,u\theta(u)\nonumber\\&+O(r^{-3}),
\label{flat space limit scalar}
\end{align}
which is exactly the expected multipole structure of a flat spacetime field: a sharp 
radiative term plus the $1/r,1/r^2$ nonradiative pieces required for $\ell\ge1$. 
The gravitational contribution ($M\neq0$) therefore encodes, on top of this, the backscattering tail $M\theta(\tilde u)$ of Eq. \eqref{scalar massless field in the corrected retarded time} and the shift $u\rightarrow \tilde u$.
\section{An overview on the massive case}
\label{The full spectrum of a massive scalar field schwarzschild}

The behavior of a massive field is radically different with respect to a massless one. 
Such a difference could be observed, from an analytical perspective, by computing 
their Green functions and noting that they cannot be related by a limiting procedure 
(cf. Sec. 5.2.1 in Ref. \cite{Esposito:2026las}). However, at least the first steps 
are equal to the previous case. Indeed, the general field expression
\begin{equation}
\psi(t,r,x^A;\mu)=\mathcal{N}
\sum_{\ell=0}^\infty\frac{     \mathbf{\Psi}_{\ell \mu}(t,r)}{r}
\sum_{m=-\ell}^\ell C_\ell^m Y_\ell^m(x^A),
\label{Schwarzschild full massive field}
\end{equation}
has the $(t,r)$ part of the field which is now given by
\begin{align}
    \psi_{\ell \mu}(t,r)=\int_{-\infty}^\infty \frac{d\omega}{\sqrt{2\pi}}&
e^{-i\omega t-r\sqrt{\mu^2-\omega^2}+\frac{M(2\omega^2-\mu^2)}
{\sqrt{\mu^2-\omega^2}}\ln r}\nonumber\\&\times\left(1+\sum_{s=1}^{\infty}
\frac{B_s(\ell,\omega;\mu)}{r^s}\right).
    \label{tr field scharzschild massive}
\end{align}
In order to write compactly the first two coefficients of the series, we introduce the quantities
\begin{align}
    L \equiv &\ell(\ell+1), \\
    \tilde\omega \equiv& \sqrt{\mu^{2}-\omega^{2}}, \label{definition of omega tilde}\\
    N \equiv& 8\omega^{4}-12\mu^{2}\omega^{2}+3\mu^{4},
\end{align}
which make it possible for us to write
\begin{align}
    B_1(\ell,\omega;\mu)=&\frac{L}{2   \tilde\omega}+\frac{M\left(2\omega^{2}
-\mu^{2}\right)}{2   \tilde\omega^{2}}+\frac{M^{2}N}{2   \tilde\omega^{3}},\\
    B_2(\ell,\omega;\mu)=&\frac{L(L-2)}{8   \tilde\omega^{2}}
  +\frac{M\left[\mu^{2}+2(L-1)\omega^{2}\right]}{4   \tilde\omega^{3}}\nonumber\\+&
  \frac{M^{2}\left[(2L-1)N+4\omega^{4}\right]}{8   \tilde\omega^{4}} \nonumber\\+&
  \frac{M^{3}\mu^{2}\left(4\omega^{4}-6\mu^{2}\omega^{2}+\mu^{4}\right)}{2   
\tilde\omega^{5}}\nonumber\\+& \frac{M^{4}N^{2}}{8   \tilde\omega^{6}}.
\end{align}
The mass signature, compared to the massless case in Eq. \eqref{Schwarzschild 
u-r massless field}, is encoded in the nonlinearity in $\omega$ of the exponent in Eq. 
\eqref{tr field scharzschild massive} and a highly nontrivial $\omega$ dependence 
of the $B_s(\ell,\omega;\mu)$ coefficient: such a feature forbids (to the best 
of our knowledge) an a priori exact evaluation of the frequency integral. 

\subsection{The leading order}
\label{subsection leading order massive scalar field}

However, several features can be extracted by considering the leading term in the 
large-$r$ expansion. In the $r\gg2M$ spacetime region we have
\begin{equation}
    F(t,r;\mu)=\int_{-\infty}^{\infty}\frac{d\omega}{\sqrt{2\pi}}
e^{-i\omega t-r\sqrt{\mu^2-\omega^2}+\frac{M(2\omega^2-\mu^2)}{\sqrt{\mu^2-\omega^2}}\ln r}.
        \label{F integral NOT rewritten fancy}
\end{equation}
This integral is the same considered in Ref. \cite{Esposito:2025jsq} (cf. Eqs (4) 
and (49)) and reconsidered in Ref. \cite{Esposito:2026las} (cf. Eqs (85)-(86) with 
$Q=a=0$), where we computed the spectrum by symmetrising the contribution of an 
ingoing and outgoing wave. In this paper we want to split the retarded and 
advanced description clearly, in order to have a more physical perspective on the 
problem. An outgoing wave is labelled by $F_R(t,r;\mu)$, where $_R$ stands for 
retarded. The advanced description is given by $F_A(t,r;\mu)=F_R(-t,r;\mu)$, 
by virtue of the time-reversal symmetry of Schwarzschild spacetime.
In Appendix ~\ref{appendix section bessel functions} we recall some useful properties 
of Bessel functions, while in Appendix ~\ref{Evaluation of the leading term of the massive scalar field} 
we describe in detail the evaluation of the integral in Eq. 
\eqref{F integral NOT rewritten fancy}. Upon introducing the functions (cf. Eqs 
\eqref{appendix definition of xi}, \eqref{appendix definition of beta})
\begin{align}
\xi(\tilde u,\tilde v)=&\sqrt{\frac{\tilde u}{\tilde v}},\\
\beta(r)=&-2M\mu\ln r,
\end{align}
where $\tilde u=u-2M\ln 2M+O(M^2/r)$ and $\tilde  v=v+2M\ln 2M+O(M^2/r)$ are properly 
introduced in the Appendix and can be understood as the corrected retarded and advanced 
Schwarzschild times, respectively. They are defined in the Appendix and differ from the 
retarded and advanced Schwarzschild time by the constant shifts $\mp2M\ln2M$.

Equation \eqref{F integral NOT rewritten fancy} reads as
\begin{equation}
F_R(\tilde u,\tilde v;\mu)=\sqrt{2\pi}(\partial_{\tilde u}-\partial_{\tilde v})
\left[\theta(\tilde u)\mathcal{J}(\tilde u,\tilde v;\mu)\right],
\label{FR u>0 massive schwarzschild}
\end{equation}
with
\begin{equation}
\mathcal{J}(\tilde u,\tilde v;\mu)=\sum_{\nu=0}^\infty c_\nu(\beta)\left(
\frac{\tilde u}{\tilde v}\right)^{\nu/2}J_\nu\left(\mu\sqrt{\tilde u\tilde v}\right),
\end{equation}
where the $c_\nu(\beta)$ coefficients are given in Eqs \eqref{Appendix c0 coefficient}, 
\eqref{appendix cnu coefficients}. The structure of a wave propagating in curved 
spacetimes described in Sec. ~\ref{The Huygens' principle and the decomposition 
of the Green function} is therefore fully-encoded in a single Lommel series 
(cf. Appendix ~\ref{appendix section bessel functions}).

More explicitly, we can write Eq. \eqref{FR u>0 massive schwarzschild} as
\begin{align}
    F_R(\tilde u,\tilde v;\mu)|_{u<0}=&0 \label{FR u<0 massive schwarzschild}\\
    F_R(\tilde u,\tilde v;\mu)|_{u\ge0}=& \sqrt{2\pi}\left[\delta(\tilde u)+\theta(\tilde u)
\mathcal{T}(\tilde u,\tilde v)\right],\label{FR u>0 massive schwarzschild with clear separation}
\end{align}
where the tail term is given by
\begin{align}
  \frac{2}{\mu}  \mathcal{T}(\tilde u,\tilde v)=&-\beta J_0+\left[\left(1+\tfrac{\beta^2}{2}\right)\xi  
-\xi^{-1}\right]  J_1\nonumber\\&-\tfrac{\beta^3}{6}\xi^2J_2+\big(\tfrac{\beta^4}{24}
-\tfrac{\beta^2}{2}\big)\xi^3J_3\nonumber\\&+\big(\tfrac{\beta^3}{3}-\tfrac{\beta^5}{120}\big)\xi^4J_4+\dots.
    \label{neumann series massive schwarzschild tail}
\end{align}
We are now able to recognise, also in the massive case, the prompt and tail 
structure. Eq. \eqref{FR u<0 massive schwarzschild} correctly describes that 
the retarded description of an outgoing wave has no support in the past 
lightcone. Eq. \eqref{FR u>0 massive schwarzschild with clear separation} gives 
instead the prompt signal through the $\delta(u)$ term and the tail governed by 
the $\mathcal{T}(\tilde u,\tilde v)$ function. The last piece is a convergent Neumann series 
supported inside the lightcone, following the prescription given in Sec. 
~\ref{The Huygens' principle and the decomposition of the Green function}.

\subsubsection{The flat spacetime limit}

The flat spacetime limit can be easily computed from Eq. 
\eqref{FR u>0 massive schwarzschild with clear separation}.
Indeed, $\beta=0$ and $u,v$ become standard Minkowski null coordinates. Therefore, 
only the $J_1$ term survives. By explicitly writing all variables, we have\footnote{Notice 
that, as a distribution in $t$ at fixed $r>0$, the Synge world function in Eq. 
\ref{synge world function in flat spacetime} vanishes at $t=\pm r$ with 
$\left|\frac{d\sigma}{dt}\right|=|t|=r$. In the retarded domain we have 
therefore $\delta(\sigma)=\delta(t-r)/r$.}
\begin{align}
    \frac{\psi_R(t,r,x^A;\mu)}{\mathcal{N}\sqrt{2\pi}}\Big|_{M=0}=&\frac{\delta(t-r)}{r} 
- \mu \theta(t-r)\frac{J_1\left(\mu\sqrt{t^2-r^2}\right)}{\sqrt{t^2-r^2}}
\nonumber\\\times&\sum_{\ell=0}^{\infty}\sum_{m=-\ell}^\ell C_\ell^m Y_\ell^m(x^A),
\end{align}
which resembles the Fourier representation of the retarded Klein-Gordon propagator 
$G_\text{ret}(x,0)$ (i.e., the Green function given in 
Eq. \eqref{scalar green function massive field flat spacetime}).
However, since we are solving the source-free equation, what the $M=0$ limit really 
confirms is that the front and the flat spacetime tail of the retarded Green 
function are reproduced correctly.

\subsubsection{The intermediate-time behavior}

It would be tempting to compute the $t\rightarrow\infty$ of the Neumann series we derived. 
However, there is an important caveat. Eq. \eqref{FR u>0 massive schwarzschild}, 
and in particular the series in Eq. \eqref{neumann series massive schwarzschild tail} 
converges since $\xi<1$ for every finite $t$ (cf. Appendix ~\ref{Evaluation of the leading term of the massive scalar field}). In the $t\rightarrow \infty$ limit, $\xi \rightarrow 1$ and 
the number of terms which should be considered in order to have an accurate estimate 
grows like $\mu\sqrt{uv}$.  Upon relying on the asymptotic expansion of Bessel functions 
(cf. Eq. \eqref{appendix asymptotic expansion bessel coefficients})
\begin{equation}
J_\nu{(Z)}\simeq \sqrt{\frac{2}{\pi Z}}\left[\cos\left(Z-\nu\frac{\pi}{2}
-\frac{\pi}{4}\right)+O(Z^{-1})\right],
\end{equation}
and introducing the variable (cf. Eq. \eqref{definition of a for the massive 
scalar schwarzschild leading integral})
\begin{equation}
    a=r+2M\ln r,
\end{equation}
Eq. \eqref{FR u>0 massive schwarzschild} becomes
\begin{align}
    F_R(t,r_*)\simeq&2a\sqrt{\mu}(t^2-a^2)^{-\frac{3}{4}}\nonumber\\
&\times\cos\left[\frac{\pi}{4}+\left(\mu+\frac{\beta }{2 a}\right) \sqrt{(t^2-a^2)} \right.\nonumber\\\
&-\left.\frac{\beta^2  t^2}{8\mu a^4} \sqrt{t^2-a^2} \right],
\end{align}
with range of validity given by
\begin{equation}
    t\ll\left(\frac{8\mu a^4}{\beta^2}\right)^\frac{1}{3} \approx8a.
\end{equation}

The last approximation is valid, e.g., for $r=10^4 M, \mu M=0.5$. Such an expression can also be derived by applying the stationary phase method 
to Eq. \eqref{F integral NOT rewritten fancy}  and provides an approximation in the range
\begin{equation}
    a\ll t\ll t_c=\frac{2}{3\sqrt{3}} \sqrt{\frac{a^3}{M\ln r}},
\end{equation}
where $t_c$ is the point where two stationary points merge (obtained from the vanishing of 
the first and second derivative of the phase). Therefore, no expansion of this kind exists beyond $t_c$.

This is just the flat massive $\ell=0$ tail, with a Coulomb phase that lowers the 
frequency by $\frac{\beta}{2r_*}=-M\mu \ln r /r_*$ for $t\gg a$, while for 
$t\sim a$ the instantaneous frequency is $\mu t/\sqrt{t^2-a^2} +\beta t/(2a\sqrt{t^2-a^2})$. 
The result we obtained can be easily compared with the one found in Ref. \cite{Hod:1998ra}, 
where the intermediate late-time behavior of a massive field is an oscillatory inverse 
power law $t^{-(\ell+3/2)}\sin(\mu t)$, valid in the Hod-Piran window $\mu M \ll \mu t \ll (\mu M)^{-2}$. 
Analytical results agree with this, as shown in Refs \cite{Koyama:2000hj,Koyama:2001ee}. We derived the $\ell=0$ case, with the same power law 
(for $t\gg a$) and oscillations governed by the field mass and an accumulating Coulomb 
phase. At $M=0$ it is exactly the flat spacetime tail of the retarded Green function. 

\section{Full spectrum of spin-weighted massless fields}
\label{Gravitational waves}

The Poincar\'e series expansion introduced in Sec. ~\ref{Poincare asymptotic expansion} 
allows the study of more involved fields in more general black hole spacetimes. 
In Refs \cite{Teukolsky:1973ha, Press:1973zz, Teukolsky:1974yv}, 
Teukolsky described massless spin-weighted perturbations of the Kerr black hole. 
The mathematical framework exploited relies on the Newman-Penrose formalism, culminating 
in the so-called Teukolsky master equation describing the radial part of the field. 
Applications to astrophysical observations are nowadays, asides the pure mathematical 
interest, of main interest. We introduced the topic and a preliminary analysis in 
Sec. VI of Ref. \cite{Esposito:2026las}. In this work we provided some basics on the 
main formulae in the Schwarzschild case. In the following we aim at a more detailed analysis of the 
full spectrum of arbitrary spin-weighted fields at null infinity and, 
in particular, of gravitational waves.

The field decomposition of Eq. \eqref{our generic field decomposition} is still 
valid, leading to a spin-weighted generalisation (which can be easily recognised 
in every formula we write) of Eq. \eqref{Schwarzschild full massless field}:
\begin{align}
    \psi^{(\mathbb{s})}(t,r,x^A;0)=&\mathcal{N}\sum_{\ell=\text{max}(|\mathbb{s}|,|m|)}^{\infty}
\frac{  \mathbf{\Psi}^{(\mathbb{s})}_{\ell 0}(t,r)}{r^{(2\mathbb{s}+1)}}
\nonumber\\\times &\sum_{m=-\ell}^{\ell}
N_{\ell m \mathbb{s}}Y^{(\mathbb{s})}_{\ell m}(x^A),
\label{general decomposition massless spin s field}
\end{align}
with $f(r)=1-2M/r$ as introduced in Eq. \eqref{1-2M on R schwarzschil factor} and 
\begin{equation}
\mathbf{\Psi}^{(\mathbb{s})}_{\ell 0}(t,r)\equiv  f(r)^{-\frac{\mathbf{s}+1}{2}}        
\psi^{(\mathbb{s})}_{\ell 0}(t,r)
\label{definition bold psi for spin}
\end{equation}
is the spin-weighted generalisation of Eq. \eqref{mathbf psi scalar}.
The angular part is described by spin-weighted spherical harmonics 
$Y^{(\mathbb{s})}_{\ell m}(x^A)$, which can be written explicitly in terms of 
Jacobi polynomials $P_{\ell-m}^{(|m+\mathbb{s}|,| m-\mathbb{s}|)}(\cos\theta)$ as
\begin{align}
    Y^{(\mathbb{s})}_{\ell m}(x^A)=&e^{im\phi}(1-\cos\theta)^\frac{|m
+\mathbb{s}|}{2}(1+\cos\theta)^\frac{|m-\mathbb{s}|}{2}\nonumber\\
\times &P_{\ell-\text{max}(|m|,|\mathbb{s}|)}^{(|m+\mathbb{s}|, |m-\mathbb{s}|)}(\cos\theta).
\end{align}

Jacobi polynomials are given by (cf. Eq. (18.5.8) in Ref. \cite{NIST:DLMF})
\begin{align}
P_k^{(|m+\mathbb{s}|,| m-\mathbb{s}|)}(\cos\theta)=\frac{1}{2^k}\sum_{i=0}^k&\binom{k
+|m+\mathbb{s}|}{i}\binom{k+|m-\mathbb{s}|}{k-i}\nonumber\\&
\times(\cos\theta-1)^{k-i}(\cos\theta+1)^{i}.
\end{align}
The separation constant is instead given by
\begin{equation}
    \lambda^{(\mathbb{s})}_{\ell}=\ell(\ell+1)-\mathbb{s}(\mathbb{s}+1)\equiv \lambda.
    \label{lambda spin weighted separation constant}
\end{equation}

The ordinary radial differential equation exhibits the coefficients
\begin{align}
    p^{(\mathbb{s})}(r)=&\frac{2(r-M)}{r^2 f(r)}{(\mathbb{s}+1)},\label{ps r}\\
    q^{(\mathbb{s})}(r)=&\frac{1}{r^2f(r)}\left[\frac{r^4\omega^2-2i{\mathbb{s}} 
\omega r^2(r-M)}{r^2f(r)}+4i\omega {\mathbb{s}} r-\lambda\right].\label{qs r}
\end{align}
Following the same route of Sec. ~\ref{FULL SPECTRUM OF A MASSLESS SCALAR 
FIELD in schwarzschild}, i.e. the introduction of the retarded branch 
$\sqrt{-\omega^2}=-i\omega$ and of the Schwarzschild retarded time $u$ in 
order to properly study an outgoing wave at future null infinity, we obtain that the integral 
underpinning the $(u,r)$ part of the field is given by
\begin{align}
    \psi_{\ell 0}^{(\mathbb{s})}(u,r)=\int_{-\infty}^\infty &\frac{d\omega}{\sqrt{2\pi}}
e^{-i\omega \left(u-2M\ln \frac{2Mr}{(r-2M)}\right)}
\nonumber\\
&\times \left(1+\sum_{k=1}^{\infty}\frac{B^{(\mathbb{s})}_k(\ell,\omega;0)}
{r^k}\right),\label{ur integral for generic s}
\end{align}
which exhibits exactly the same structure of Eq. \eqref{Schwarzschild u-r massless field}, 
since the spin dependence enters only throughout the $B^{(\mathbb{s})}_s(\ell,\omega;0)$ 
coefficients. The first of them reads as\footnote{Notice the typo in Eq. 
(139) of Ref.\cite{Esposito:2026las}: $\lambda^2$ should read simply as $\lambda$. Therefore, 
one should keep the square root of $\lambda$ in Eqs (143)-(145), (147), (148).} 
\begin{align}
    B^{(\mathbb{s})}_1(\ell,\omega;0)=&\frac{i\lambda}{2\omega}-4i\omega M^2-M(\mathbb{s}+1), \label{b1 coefficient s}\\
    B^{(\mathbb{s})}_2(\ell,\omega;0)=&-8M^4\omega^2+4iM^3\omega 
\mathbb{s}\nonumber \\+&\frac{M^2}{2}(4\lambda+\mathbb{s}^2-1)
\nonumber \\+&\frac{iM}{2\omega}(\mathbb{s}+1)(2\mathbb{s}+1-\lambda)
\nonumber\\+&\frac{\lambda}{8\omega^2}(2\mathbb{s}+2-\lambda).\label{b2 coefficient s}
\end{align}
Such Laurent polynomials lead 
to the fundamental integral given in Eq. \eqref{the massless In integral}, whose 
solutions are provided by Eq. \eqref{evaluation of the massless In integral}. The 
mathematical and physical meaning of each term is therefore clear. Indeed, Eq. 
\eqref{Schwarzschild u-r massless field explicit up to r-2} reads now
\begin{align}
    \frac{\psi^{(\mathbb{s})}_{\ell 0}(u,r)}{\sqrt{2\pi}}=&\varrho^{(\mathbb{s})}_0(r)
\delta(A)+\varrho^{(\mathbb{s})}_1(r)\delta^{(1)}(A)\nonumber\\+&\varrho_2(r)\delta^{(2)}(A)
+2\varsigma^{(\mathbb{s})}_1(r)\theta(A)\nonumber\\+&2\varsigma^{(\mathbb{s})}_2(r)A\theta(A)+O(r^{-3}),
\label{massless field arbitrary spin u r part in A}
\end{align}
with the following form of the regular $\varrho_n^{(\mathbb{s})}$ 
and singular $\varsigma_n^{(\mathbb{s})}$ coefficients:
\begin{align}
\varrho^{(\mathbb{s})}_0=&1-\frac{(\mathbb{s}+1)M}{r}+\frac{M^2(4\lambda+\mathbb{s}^2-1)}{2r^2},\\
\varrho^{(\mathbb{s})}_1=&\frac{4M^2}{r}-\frac{4\mathbb{s}M^3}{r^2},\\
\varrho_2=&\frac{8M^4}{r^2},\\
\varsigma^{(\mathbb{s})}_1=&\frac{\lambda}{4r}+\frac{M(\mathbb{s}+1)(2\mathbb{s}+1-\lambda)}{4r^2},\\
\varsigma^{(\mathbb{s})}_2=&\frac{(\ell-\mathbb{s})(\ell-\mathbb{s}-1)(\ell+\mathbb{s}+1)(\ell+\mathbb{s}+2)}{16r^2}.
\end{align}
A first consistency check is given by the $\mathbb{s}=0$ case, where each term reproduces 
the scalar scenario (since $\lambda_{\mathbb{s}=0}=\ell(\ell+1)$ there). It is interesting to write the result also with respect to the 
corrected retarded time $\tilde u$ defined in Eq. \eqref{definition of the corrected retarded time}, 
resembling a structure very similar to Eq. \eqref{scalar massless field in the corrected 
retarded time}, i.e., its spin-weighted generalisation
\begin{align}
 \frac{  \mathbf{\Psi}^{(\mathbb{s})}_{\ell 0}(u,r)}{\sqrt{2\pi}}=&\delta(\tilde u)
+\left[\frac{\lambda}{2r}+\frac{M}{2r^2}(\mathbb{s}+1)(2\mathbb{s}+1)\right]\theta(\tilde u)
\nonumber\\+&\frac{1}{8r^2}(\ell-\mathbb{s})(\ell-\mathbb{s}-1)(\ell
+\mathbb{s}+1)\nonumber\\\times&(\ell+\mathbb{s}+2)\tilde u\theta(\tilde u)\nonumber\\+&O(r^{-3}).
\label{massless field arbitrary spin u r part in utilde}
\end{align}
The most remarkable feature is that the prompt signal $\delta(\tilde u)$ is spin-independent. 
Moreover, the $M/r^2$ correction to $\theta(\tilde u)$ is $\ell$-independent and vanishes 
only for $\mathbb{s}\in\left\{-\frac{1}{2},-1\right\}$. The last term before the remainder 
vanishes instead for $\ell=\{\mathbb{s},\mathbb{s}+1\}$, 
recovering the scalar value only for $\mathbb{s}=-2$.

\section{Results and concluding remarks}

Now we find it appropriate to begin a detailed assessment of our results,
divided into three parts.

\subsection{A brief comparison with the literature}
\label{sec: a brief comparison with the literature}

Equation \eqref{massless field arbitrary spin u r part in utilde} can be understood as 
the main result of this paper, presented as a spin-weighted generalisation of the 
formula obtained for massless scalar fields. It is therefore of great interest to 
contextualise it with respect to what is already known on the topic. We underpin 
that Eq. \eqref{massless field arbitrary spin u r part in utilde} is a large-$r$ 
expansion at fixed retarded time, towards future null infinity. Price-law papers 
connect to it only through the $1/r$ hierarchy
at null infinity. Furthermore, it is written as distributions, assuming a unit spectral 
weight. For this reason our result behaves as a Green function. Papers that start 
from initial data or sources carry extra frequency-dependent amplitudes.

In his seminal paper on black hole perturbation theory, Teukolsky (cf. Eqs (5.3), 
(5.4) of Ref. \cite{Teukolsky:1973ha}) obtained that the asymptotic radial part of 
a field in Kerr spacetime (where the angular black hole velocity enters just as a 
mere parameter) describing an outgoing wave is given by $e^{i\omega r_*}/r^{2\mathbb{s}+1}$, 
which is exactly reproduced by the fall-off behavior of Eq. \eqref{general decomposition 
massless spin s field}. Another comparison with Teukolsky's work has to do with the relation 
among opposite spins of the same radial part. Indeed, 
Eq. (3.22) of Ref. \cite{Teukolsky:1974yv} reads as
\begin{equation}
\mathcal{D}\mathcal{D}\mathcal{D}\mathcal{D}R_{-2}=\frac{1}{4}R_{2},
\end{equation}
which gives the relation among the opposite spins of the radial equation describing 
gravitational waves. 
The $1/4$ factor refers to the normalisation of Ref. \cite{Teukolsky:1974yv}.
The operator $\mathcal{D}$ is given by
\begin{align}
\mathcal{D}=&\partial_r -i\frac{\omega}{f(r)},
\end{align}
while $R_{\mathbb{s}}$ is given by the $r$-dependent part of Eq. \eqref{ur integral 
for generic s}, i.e., 
\begin{equation}
    R_{\mathbb s}(\omega,r)=r^{-(2\mathbb s+1)}f^{-\frac{\mathbb s+1}{2}}\,e^{i\omega(r+2M\ln r)}\left(1+\sum_{k\ge1}\frac{B^{(\mathbb s)}_k}{r^k}\right).
\end{equation}
In our case, with both radial functions normalised as in Eq. \eqref{ur integral for generic s} (i.e., $B_0^{(\mathbb{s})}=1$),  Teukolsky's relation reads
\begin{equation}
\mathcal{D}\mathcal{D}\mathcal{D}\mathcal{D}R_{-2}=\frac{|\Lambda(\Lambda+2)-12iM\omega|^2}{16\omega^4}R_{2},
\label{relation s 2 to s -2}
\end{equation}
with $\Lambda\equiv\lambda^{-2}=(\ell-1)(\ell+2)$, i.e., the two components are related by a frequency 
dependent factor. This relation needs more clarifications. The differential operator 
$\mathcal{D}^4$ removes the first four terms of the $\mathbb{s}=-2$ series (from $r^3$ 
to $r^0)$. Therefore, the $1/r^k$ correction of the $\mathbb{s}=2$ field is matched by 
the $B^{(-2)}_{k+3}$ and $B^{(-2)}_{k+4}$ of the $\mathbb{s}=-2$ series, plus all the subleading terms. 
We therefore have computed the identity at subleading orders of the series expansion, 
with some of them shown in the Appendix. This means that at unit spectral weight, the 
two $\mathbb{s}=\pm 2$ components come from different sources or, the perturbation 
generating the $\mathbb{s}=2$ component is related with the one generating the 
$\mathbb{s}=-2$ component through the frequency dependent factor in Eq. \eqref{relation s 2 to s -2}.

Mathematical properties of the Teukolsky equation, whose radial part in our case reduces 
to Eq. \eqref{second order radial equation in the right form for poincare} with 
coefficients given by Eqs \eqref{ps r}, \eqref{qs r}, are deeply investigated by 
Leaver in Ref. \cite{Leaver:1986vnb}. In particular, in his Eq. (122) he gives the asymptotic 
expansion of the radial equation in terms of irregular Coulomb wave-functions (providing a 
convergence criterium), which resembles the structure of ours. More comments on such functions, 
followed by a $2M\omega$ (a post-Newtonian inspired) expansion, are given in Ref. 
\cite{Mano:1996vt} and in the review in Ref. \cite{Sasaki:2003xr}.

As already pointed out by us in Ref. \cite{Esposito:2025jsq}, the Poincar\'e asymptotic 
expansion through Eq. \eqref{canonical form of the radial equation at infinity} 
makes it possible for us to go beyond the Whittaker approximation used in the literature, 
starting from, e.g., Ref. \cite{Rowan:1977zg}. Indeed, the $B_1^{(\mathbb{s})}(\ell,\omega;0)$ 
coefficient in Eq. \eqref{b1 coefficient s} equals the first Whittaker (i.e., Coulomb) 
coefficient exactly. $B_2^{(\mathbb{s})}(\ell,\omega;0)$ in Eq. \eqref{b2 coefficient s}, 
instead, differs from the Whittaker one by
\begin{align}
B_2^{(\mathbb{s})}(\ell,\omega;0)-B^W_2=&\frac{A_3}{4i\omega} \nonumber\\=&-8iM^3\omega+\frac{iML}{2\omega}.
\end{align}
The purity of the prompt signal at $O(M^3/r^2)$ and the $\ell$-independence of the 
backscattering step are beyond-Coulomb effects, encoded by the $A_3$ coefficient. This 
is a clear contact with the works based on the Coulomb function we cited. We are able 
to describe qualitative different features thanks to a more accurate expansion, leading 
however to a priori divergent Poincar\'e series. This is not a problem, provided that 
observations are made in a suitable asymptotic region with respect to the source. 
A clear connection with retarded multipole solutions of the wave equation in post-Newtonian 
gravity  is given by considering Ref. \cite{Blanchet:2013haa}: the $M=0$ version of Eq. 
\eqref{massless field arbitrary spin u r part in utilde} resembles the retarded multipole 
structure, giving in the $\mathbb{s}=0$ case the $a_1=L/2$ and $a_2=L(L-2)/8$ 
coefficients of $h_\ell^{(1)}$.

\subsection{Plots}
\label{section plots}

\begin{figure*}[t]
\centering
\includegraphics[width=\textwidth]{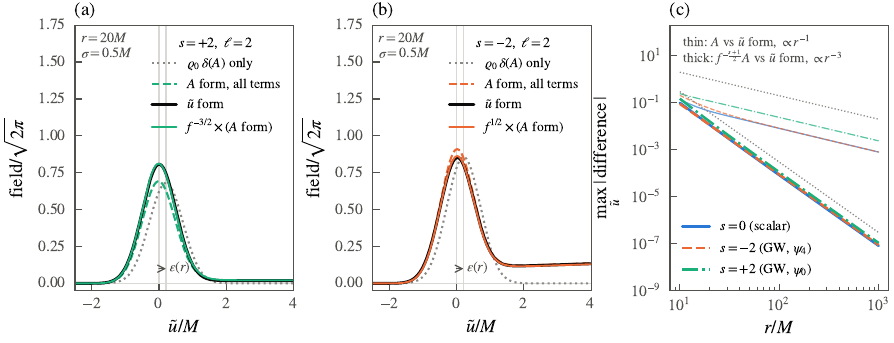}
\caption{The $A$ form [Eq.~\eqref{massless field arbitrary spin u r part in A}] and the $\tilde u$ form 
[Eq.~\eqref{massless field arbitrary spin u r part in utilde}] describe the same field. (a) $\mathbb{s}=+2$, 
(b) $\mathbb{s}=-2$, with $\ell=2$, $r=20M$ and $\sigma=0.5M$: the $\varrho_0\,\delta(A)$ term alone 
(dotted) is displaced by $\varepsilon(r)$; the full $A$ form (dashed) is centred again at $\tilde u=0$, and 
multiplying it by $f(r)^{-\frac{\mathbb{s}+1}{2}}$ (thin solid) reproduces the $\tilde u$ form (thick black). 
(c) Maximum difference over $\tilde u$ between the two forms, as a function of $r$: $\propto r^{-1}$ 
without the factor $f(r)^{-\frac{\mathbb{s}+1}{2}}$ (thin lines), $\propto r^{-3}$ with it (thick lines).}
\label{fig:spin frames}
\end{figure*}

The results of Sec.~\ref{Gravitational waves} are distributions, i.e., generalized quantity 
computed in the abstract case of unit spectral weight. In order to recover physical 
predictions, we must therefore specify the nature of the spectral emission of the source we 
are considering. A convenient and very general choice is given by folding 
Eq.~\eqref{massless field arbitrary spin u r part in utilde} with a Gaussian pulse of duration $\sigma$, 
i.e., with the source spectrum (we highlight that upon changing such a function, 
one can describe mathematically every possible source)
\begin{equation}
    \mathbb{g}(\omega)=e^{-        \frac{\sigma^2\omega^2}{2}}.
\end{equation}
Every distribution is then replaced by a smooth function,
\begin{align}
\delta(\tilde u)\to&\frac{e^{-\tilde u^2/2\sigma^2}}{\sigma\sqrt{2\pi}},\qquad
\theta(\tilde u)\to\Phi\!\left(\frac{\tilde u}{\sigma}\right),\nonumber\\
\tilde u\,\theta(\tilde u)\to&\,\sigma\left[x\,\Phi(x)+\phi(x)\right]_{x=\tilde u/\sigma},
\label{gaussian smearing}
\end{align}
where $\phi$ and $\Phi$ are the standard normal density and its cumulative distribution function. 
The plotted quantity is denoted, for convenience and readability, by the notation 
\begin{equation}
    X\equiv  \frac{ \mathbf{\Psi}^{(\mathbb{s})}_{\ell 0}(t,r)}{\sqrt{2\pi}},
    \label{quantity X}
\end{equation} 
i.e., $r^{2\mathbb{s}+1}\psi^{(\mathbb{s})}$ up to the normalization and the angular factor 
(cf.\ Eq.~\eqref{general decomposition massless spin s field}), so that fields with different 
spin weight can be compared on the same scale. We give special emphasis to the scalar field 
($\mathbb{s}=0$) and to the two gravitational components $\mathbb{s}=-2$ and $\mathbb{s}=+2$, i.e.\ the 
Weyl scalars $\psi_4$ and $\psi_0$ (cf. Ref. \cite{Teukolsky:1973ha}). Since 
Eq.~\eqref{massless field arbitrary spin u r part in utilde} is truncated at $O(r^{-2})$, it is reliable 
only for $\tilde u\ll r$: in every window shown (apart Fig. \eqref{fig:spin frames} (a)), 
the next order of the expansion stays below $15\%$ 
of the tail.
 
\begin{figure*}[t]
\centering
\includegraphics[width=\textwidth]{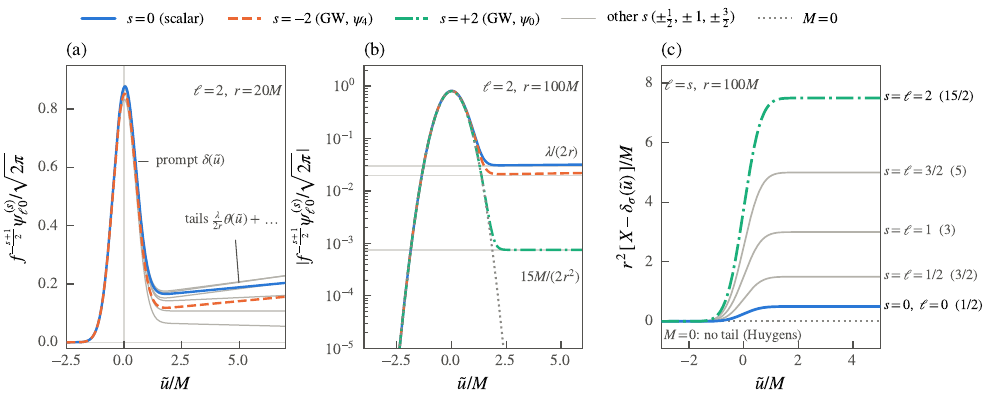}
\caption{Prompt and tail of spin-weighted massless fields far from a Schwarzschild black hole, 
Eq.~\eqref{massless field arbitrary spin u r part in utilde} folded with a Gaussian pulse of duration 
$\sigma=0.5M$ [Eq.~\eqref{gaussian smearing}]; the spin weight $\mathbb{s}$ is labelled $s$ in the plots. 
(a) $\ell=2$, $r=20M$: the prompt $\delta(\tilde u)$ is the same for every spin weight; the tails switch 
on at $\tilde u=0$ and grow linearly, with the same slope for the scalar field and for $\psi_4$ 
($\mathbb{s}=-2$). Grey lines: $\mathbb{s}=\pm\frac12,\pm1,\pm\frac32$. (b) $\ell=2$, $r=100M$, 
logarithmic scale: for $\psi_0$ ($\mathbb{s}=+2$, $\lambda=0$) the flat-space field obeys Huygens' 
principle (dotted, $M=0$) and the whole tail, $15M/(2r^2)$, is due to the curvature. 
(c) Curvature-induced tails for $\ell=\mathbb{s}$, $r=100M$, rescaled by $r^2/M$: their height 
$(\mathbb{s}+1)(2\mathbb{s}+1)/2$ grows from $1/2$ for the scalar monopole to $15/2$ for the 
gravitational quadrupole.}
\label{fig:spin prompt tail}
\end{figure*}

Figure~\ref{fig:spin prompt tail} displays the prompt and tail structure. In panel (a) the prompt 
pulse is the same for all spin weights, as stated below 
Eq.~\eqref{massless field arbitrary spin u r part in utilde}, while the tails switch on at the light 
front $\tilde u=0$ with height $\lambda/(2r)$ plus the curvature correction, and then grow linearly. 
The linear growth of the $\mathbb{s}=-2$ field is parallel to the scalar one, since the coefficient of 
$\tilde u\,\theta(\tilde u)$ coincides for $\mathbb{s}=0$ and $\mathbb{s}=-2$. Panel (b) shows the same 
structure on a logarithmic scale and highlights the gravitational field $\psi_0$ with $\ell=2$: since 
$\lambda=0$ for $\ell=\mathbb{s}$, in flat spacetime this field obeys Huygens' principle exactly, and 
its whole tail, of height $15M/(2r^2)$, is generated by the spacetime curvature. Panel (c) collects these 
purely curvature-induced tails for $\ell=\mathbb{s}$: their height $(\mathbb{s}+1)(2\mathbb{s}+1)M/(2r^2)$ 
grows with the spin weight, from $M/(2r^2)$ for the scalar monopole to $15M/(2r^2)$ for the gravitational 
quadrupole.
 
\begin{figure*}[t]
\centering
\includegraphics[width=\textwidth]{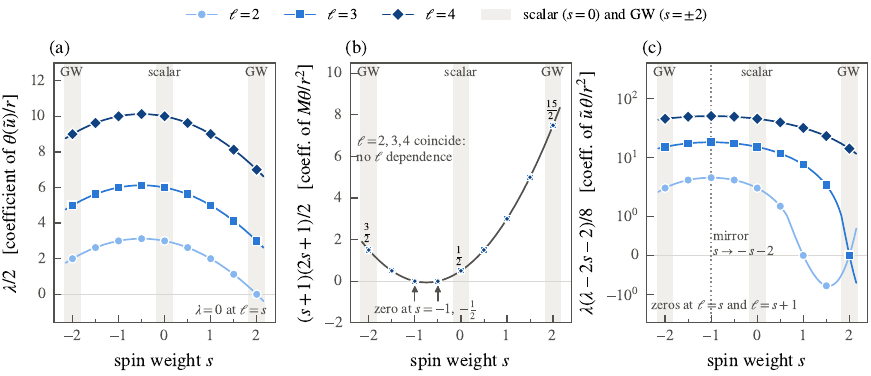}
\caption{The three tail coefficients of Eq.~\eqref{massless field arbitrary spin u r part in utilde} as 
functions of the spin weight $\mathbb{s}$, for $\ell=2,3,4$. (a) Flat-space step $\lambda/2$, which vanishes 
for $\ell=\mathbb{s}$. (b) Curvature step $(\mathbb{s}+1)(2\mathbb{s}+1)/2$: identical for every $\ell$ 
and vanishing only for $\mathbb{s}=-1,-\frac{1}{2}$. (c) Coefficient of $\tilde u\,\theta(\tilde u)$, 
$(\ell-\mathbb{s})(\ell-\mathbb{s}-1)(\ell+\mathbb{s}+1)(\ell+\mathbb{s}+2)/8$: symmetric under 
$\mathbb{s}\to-\mathbb{s}-2$ (dotted line), so that $\mathbb{s}=-2$ coincides with the scalar value, and 
vanishing for $\ell=\mathbb{s},\mathbb{s}+1$. Shaded bands: scalar and gravitational cases. Lines between 
the markers only guide the eye.}
\label{fig:spin coefficients}
\end{figure*}

Figure~\ref{fig:spin coefficients} isolates the three tail coefficients of 
Eq.~\eqref{massless field arbitrary spin u r part in utilde} as functions of $\mathbb{s}$, and makes visible 
the properties listed below it: the flat-space step $\lambda/2$ vanishes for $\ell=\mathbb{s}$; the 
curvature step $(\mathbb{s}+1)(2\mathbb{s}+1)/2$ is the same for every $\ell$ and vanishes only for 
$\mathbb{s}\in\{-1,-\frac{1}{2}\}$; the coefficient of the linear term is symmetric under 
$\mathbb{s}\to-\mathbb{s}-2$, which is why $\mathbb{s}=-2$ reproduces the scalar value, and vanishes for 
$\ell\in\{\mathbb{s},\mathbb{s}+1\}$.
 
Eventually, Fig.~\ref{fig:spin frames} compares the two forms of the same field, 
Eq.~\eqref{massless field arbitrary spin u r part in A} in terms of $A$ and 
Eq.~\eqref{massless field arbitrary spin u r part in utilde} in terms of $\tilde u$. The 
$\varrho_0\,\delta(A)$ term alone is displaced from the light front by 
$\varepsilon(r)=-2M\ln f(r)$; the 
$\varrho_1\delta^{(1)}(A)$ and $\varrho_2\delta^{(2)}(A)$ terms move it back to $\tilde u=0$, as 
anticipated by Eq.~\eqref{massless delta utilde in function of delta A}, 
and the factor $f(r)^{-\frac{\mathbb{s}+1}{2}}$ of 
Eq.~\eqref{definition bold psi for spin} restores unit amplitude. The difference between 
the two forms decreases as $r^{-3}$, consistently with the truncation. We notice that the $A$ form 
requires $\sigma\gtrsim\varepsilon(r)\simeq 4M^2/r$, since the derivatives of the Dirac delta come from 
a Taylor expansion in $\varepsilon$, while the $\tilde u$ form holds for every pulse duration.
 
\subsection{Observations and phenomenology}
\label{section observations}

For gravitational waves the radiative component is $\mathbb{s}=-2$: with the Kinnersley tetrad 
$\psi_4=\rho^4\psi^{(-2)}$, $\rho=-1/r$, so that $\psi_4\sim r^{-1}$ while $\psi_0\sim r^{-5}$ (cf. Ref.
\cite{Teukolsky:1973ha}), and far from the source $\psi_4$ is proportional to 
$\ddot h_+-i\ddot h_\times$, the second time derivative of the strain measured by ground-based 
interferometers (cf. Ref.\cite{LIGOScientific:2016aoc}). Two features of 
Eq.~\eqref{massless field arbitrary spin u r part in utilde} bear directly on these observations. 
First, the prompt signal $\delta(\tilde u)$ does not depend on the spin weight and propagates along the 
exact null cone $\tilde u=0$, which already contains the logarithmic Shapiro delay $2M\ln r$ of the 
tortoise coordinate (cf. Ref. \cite{Shapiro:1964uw}). Gravitational and electromagnetic signals emitted together 
near a black hole therefore arrive together (cf. Eq. \eqref{massless field arbitrary 
spin u r part in utilde}) : this is the property tested by the joint detection of 
GW170817 and GRB 170817A, which constrains the difference between the speed of gravitational waves and the 
speed of light to lie between $-3\times10^{-15}$ and $+7\times10^{-16}$ times the latter, and also 
constrains the difference between their Shapiro delays (cf. Ref. \cite{LIGOScientific:2017zic}). Second, for a 
signal of duration $T$ the tails are suppressed with respect to the prompt by $\lambda T/(2r)$ and 
$(\mathbb{s}+1)(2\mathbb{s}+1)MT/(2r^2)$. For GW150914, at a luminosity distance of about $410$ Mpc (cf. Ref. 
\cite{LIGOScientific:2016aoc}), $T/r\lesssim10^{-16}$: at the detector the waveform is the prompt signal, 
and the tails of Figs. \ref{fig:spin prompt tail} and \ref{fig:spin coefficients} matter in the region 
$r\sim10$--$100M$, where the signal is shaped before reaching the observer.

\subsection{Possible developments}

As far as we can see, the perspectives resulting from subsections A, B and C are
quite encouraging: our original theoretical formulae lead to predictions which can be verified
on observational ground. In the current literature, what is often studied is the single 
Fourier mode of the spectral decomposition of a field, leaving aside the whole wave packet. 
However, the latter is precisely the true physical quantity that should be considered,
and we have developed original calculations for its evaluation to the accuracy needed. 
A hard technical step might be the development of an
analogous analysis for the merging of black holes. 

\begin{acknowledgments}
M.R. acknowledges support from the INFN
Iniziativa Specifica QUAGRAP and from the European COST actions BridgeQG CA23130
and CaLISTA CA21109. G.E. is grateful to the INFN projects ET and QGSKY, and to
INDAM, for association.\\
M.R. dedicates this paper to his late father Claudio.
\end{acknowledgments}

\appendix

\section{Some remarks on Bessel functions}
\label{appendix section bessel functions}

The investigation of the full spectrum of a massive scalar field carried out 
in this work relies mainly on the mathematical theory of Bessel functions. 
In this section we aim at giving a brief overview on the main formulae considered. 
Standard references are given in Ref \cite{B1945}, Chapter VII in Ref. 
\cite{Erdelyi:1953hft}, Chapter X in Ref. \cite{NIST:DLMF}.
Bessel functions of the first kind $J_\nu(z)$ are solutions of the 
Bessel ordinary differential equation
\begin{equation}
    \left[z\frac{d}{dz}\left(z\frac{d}{dz}\right)+(z^2-\nu^2)\right]J_\nu(z)=0.
\end{equation}
This is a limiting case of the hypergeometric differential equation. It has a 
singularity of the regular type at $z=0$ and an irregular singularity at $z=\infty$. 
All other points are regular. We are interested in the special case
\begin{equation}
    \nu\in \mathbb{N}.
\end{equation}
Upon considering such a condition, $J_\nu (z)$ are known as Bessel coefficients. 
The first remarkable property is that
\begin{equation}
    J_{-\nu}(z)=(-1)^{\nu}J_\nu(z).
    \label{appendix J-nu=something J-nu}
\end{equation}
Bessel coefficients can be related to a generating function. In fact,
\begin{equation}
    e^{\frac{z}{2}(t-t^{-1})}=\sum_{\nu=-\infty}^\infty t^\nu J_\nu(z),
\end{equation}
or, more generally,
\begin{align}
     e^{\frac{z}{2}(t-a^2t^{-1})}=&\sum_{\nu=-\infty}^\infty
\left( \frac{t}{a}\right)^\nu J_\nu(az)\\&=J_0(az)+\sum_{\nu=1}^\infty 
J_\nu(az)\nonumber\\&\times\left[\left(\frac{t}{a}\right)^\nu+(-1)^\nu
\left(\frac{t}{a}\right)^{-\nu}\right]\label{appendix generating function Bessel coefficients}.
\end{align}
Of main importance for our purposes is the integral representation of Bessel 
coefficients. By applying the Cauchy theorem of residues to Eq. \eqref{appendix 
generating function Bessel coefficients} we have
\begin{equation}
    J_\nu(az)=\frac{a^\nu}{2\pi i}\int_C dt  \frac{  e^{\frac{z}{2}(t-a^2t^{-1})}}{t^{\nu+1}},
\end{equation}
where $C$ is a simple closed contour around the origin in the $t$-plane. Upon 
considering $a=1$ and choosing $C$ as the unit circle around the origin (i.e., 
$t=e^{i\phi}$), we have the Bessel representation (valid only for $\nu\in\mathbb{Z}$)
\begin{equation}
    J_\nu(z)=\frac{1}{2\pi}\int_{0}^{2\pi}d\phi e^{i(z\sin\phi -\nu \phi)}.
    \label{appendix bessel representation of bessel coefficients}
\end{equation}
In the $z\rightarrow \infty$, $\nu$ fixed regime, the following asymptotic expansion 
for Bessel functions holds:
\begin{align}
    J_\nu(z)\sim \sqrt{\frac{2}{\pi z}}&\left[\cos \omega(z) \sum_{k=0}^\infty 
(-1)^k \frac{a_{2k}(\nu)}{z^{2k}}\right.\nonumber\\-&\left.\sin \omega(z) 
\sum_{k=0}^\infty (-1)^k \frac{a_{2k+1}(\nu)}{z^{2k+1}}\right],
    \label{appendix asymptotic expansion bessel coefficients}
\end{align}
where
\begin{align}
    \omega(z)=&z-\frac{1}{2}\nu \pi -\frac{1}{4}\pi,
    \end{align}
    and the coefficients are
    \begin{align}
    a_0(\nu)=&1,\\
    a_k(\nu)=&\frac{1}{k! 8^k}(4\nu^2-1^2)(4\nu^2-3^2)\cdots[4\nu^2-(2k-1)^2] \nonumber
    \\=&\frac{1}{(-2)^k k!}\left(\frac{1}{2}-\nu\right)_k\left(\frac{1}{2}+\nu\right)_k \quad k\ge1.
\end{align}
The branch of $\sqrt{z}$ is determined by
\begin{equation}
    \sqrt{z}=e^{\frac{1}{2}(\ln |z|+i\text{ph}(z)) }.
\end{equation}
After having assumed that $\nu\ge0$, $z>0$, the remainder concerning the truncation 
of the series in Eq. \eqref{appendix asymptotic expansion bessel coefficients} 
$\sum_{k=0}^{\ell-1}(-1)^k a_{2k}(\nu)z^{-2k}$
does not exceed the first neglected term in absolute value and it exhibits 
the same sign provided that $\ell\ge \text{max}\left(\frac{1}{2}\nu-\frac{1}{4},1\right)$.  
The same holds for the odd term $ \sum_{k=0}^{\ell-1}(-1)^k a_{2k+1}(\nu)z^{-2k-1}$, 
provided that $\ell\ge \text{max}\left(\frac{1}{2}\nu-\frac{3}{4},1\right)$. 
Instead, if $\nu\rightarrow \infty$ through positive real values, 
at $z\neq 0$ fixed, we have that Bessel functions read as
\begin{equation}
    J_\nu(z)\sim \frac{1}{\sqrt{2\pi \nu}}\left(\frac{ez}{2\nu}\right)^\nu,
\end{equation}
where $e$ is the base of the natural logarithm. 

It is well known that analytic functions can be expanded in series of polynomials 
by the theorems of Taylor and Laurent. The same happens in terms of Bessel functions. 
Indeed, any series expansion of the form
\begin{equation}
    f(z)=\sum_{\nu=0}^\infty a_\nu J_{\nu}(z)
    \label{Appendix neumann expansion}
\end{equation}
is called a Neumann expansion. A Neumann series is defined as
\begin{equation}
    \sum_{n=0}^\infty a_n J_{\nu+n}(z).
\end{equation}
It can be shown that the Neumann series converges throughout the domain in which 
\begin{equation}
    \lim_{n\rightarrow\infty}\text{sup}|a_n J_{\nu+n}(z)|^{\frac{1}{n}}<1,
\end{equation}
which is identical to the domain identified by
\begin{equation}
\lim_{n\rightarrow\infty}\text{sup}\left|\frac{a_n (z/2)^{\nu+n}}{\Gamma(\nu+n+1)}\right|^{\frac{1}{n}}<1.
\end{equation}
Such a result proves that the circles of convergence of a Neumann series and 
of the associated power series are identical.
An interesting application of Neumann series is given by Lommel, who investigated 
functions of two independent variables. In particular, we are interested in the Lommel series
\begin{equation}
    U_\nu (u,v)=\sum_{m=0}^\infty(-1)^m\left(\frac{u}{v}\right)^{\nu+2m}J_{\nu+2m}(v).
    \label{appendix Lommel series}
\end{equation}
In the special case of $u=v$ we have
\begin{align}
    U_0(z,z)=&\frac{1}{2}[J_0(z)+\cos z],\\
    U_1(z,z)=&\frac{1}{2}\sin z,
\end{align}
and therefore the general expressions (here $\epsilon_{m}$ is the Neumann factor)
\begin{align}
U_{2\nu}(z,z)=&\frac{(-1)^\nu}{2}\left[\cos z
-\sum_{m=0}^{\nu-1}(-1)^m \epsilon_{2m}J_{2m}(z)\right],\\
U_{2\nu+1}(z,z)=&\frac{(-1)^\nu}{2}\left[\sin z
-\sum_{m=0}^{\nu-1}(-1)^m \epsilon_{2m+1}J_{2m+1}(z)\right],
\end{align}
provided that $\nu \ge1$, $\nu\ge0$, respectively.

Two useful asymptotic expansions for large values of $|u|$ are
\begin{align}
U_\nu (u,0)\sim &\cos\left[\frac{1}{2}\left(u-\nu \pi\right)\right]
\nonumber\\&+\sum_{m=0}^\infty \frac{(-1)^m}{\Gamma(\nu-1-2m)(u/2)^{2m-\nu+2}},\\
     U_\nu (u,v)\sim &\cos\left[\frac{1}{2}\left(u+\frac{v^2}{u}
-\nu \pi\right)\right]\nonumber\\&+\sum_{m=0}^\infty (-1)^m
\left(\frac{v}{u}\right)^{2m-\nu+2}J_{\nu-2-2m}(v).
\end{align}
 
\section{Evaluation of the generating integral of the massless scalar field}
\label{appendix section massive leading order integral}

The massless case provides a straightforward integration over 
frequencies thanks to the generating integral 
\begin{equation}
I_n=\int_{-\infty}^\infty \frac{d\omega}{\sqrt{2\pi}}
e^{-i\omega A(r)} \omega^n, \quad n\in \mathbb Z,
\end{equation}
where every $I_n$ is a distribution in $A(r)$.
The key property is that the exponent is linear in $\omega$. Indeed, the 
general form for $n\ge 0$ can be recovered from the $n=0$ case, 
which is nothing but the well-known Fourier transform
\begin{equation}
\int_{-\infty}^\infty \frac{d\omega}{\sqrt{2\pi}}
e^{-i\omega A(r)}=\sqrt{2\pi} \delta[A(r)],
\end{equation}
by taking $n$ derivatives, since
\begin{equation}
    \omega^n e^{-i\omega A(r)}=i^n \partial^{(n)}_Ae^{-i\omega A(r)},
\end{equation}
and therefore
\begin{align}
\int_{-\infty}^\infty d\omega \omega^n e^{-i\omega A(r)}
=&i^n\partial^{(n)}_A\int_{-\infty}^\infty d\omega  
e^{-i\omega A(r)}\nonumber\\=&i^n 2\pi \delta^{(n)}[A(r)].
\end{align}
The $n<0$ case relies instead on the standard result (upon considering the retarded 
choice, analytic in $\text{Im}(\omega)>0$)
\begin{equation}
\int_{-\infty}^\infty \frac{d\omega}{\sqrt{2\pi}}
\frac{e^{-i\omega A(r)}}{\omega +i0}=-i\sqrt{2\pi}\theta(A),
\end{equation}
leading to the closed formula
\begin{equation}
\int_{-\infty}^{\infty}\frac{d\omega}{\sqrt 2\pi}e^{-i\omega A(r)}(\omega+i0)^n
= i^n\sqrt{2\pi}\frac{\theta(A)}{(|n|-1)!}A^{|n|-1}(r).
\end{equation}
The first five terms are 
\begin{align}
I_0 &= \sqrt{2\pi}\,\delta(A), &\\
I_1 &= i\sqrt{2\pi}\,\delta^{(1)}(A), &\\
I_2 &= -\sqrt{2\pi}\,\delta^{(2)}(A),\\
I_{-1} &= -i\sqrt{2\pi}\,\theta(A), &\\
I_{-2} &= -\sqrt{2\pi}\,A\theta(A) . & &
\end{align}
Each of these contributions allows us to reconstruct the form of the field, as 
explained in the main part of the paper. For $n\ge0$, the $\delta^{(n)}(A)$ terms 
are the Taylor series of the shift $A\rightarrow \tilde u$ (cf. Eq. \eqref{massless delta utilde in function of delta A}) and cancel in the $\tilde u$ form of the field. 
Besides these pieces, the $n<0$ part of the 
spectrum concerns the following: increasingly infrared singular powers 
$\omega^{-|n|}$ Fourier-transform into increasingly slowly-decaying 
(here, growing) functions of $A$, the seed of late-time power-law tails. 

\section{Evaluation of the leading term of the massive scalar field}
\label{Evaluation of the leading term of the massive scalar field}
Since it is a slightly lengthy computation, we report here the evaluation 
of the integral in Eq. \eqref{F integral NOT rewritten fancy}, 
which can be rewritten as
\begin{equation}
F(t,r;\mu)=\int_{-\infty}^{\infty}\frac{d\omega}{\sqrt{2\pi}}
e^{-i\omega t-a\tilde \omega -\frac{b}{\tilde \omega}},
\label{F integral rewritten fancy}
\end{equation}
where 
\begin{equation}
a=r+2M\ln r,
\label{definition of a for the massive scalar schwarzschild leading integral}
\end{equation}
can be understood (as we underpin in the main text) as the large-$r$ 
expansion of the tortoise coordinate in Eq. 
\eqref{Schwarzschild tortoise coordinate} and
\begin{equation}
    b=-M\mu^2\ln r
\end{equation}
encodes the logarithmic tail that results from the coupling of the 
field mass $\mu$ with the spacetime curvature. 
The function $\tilde \omega=\sqrt{\mu^2-\omega^2}$ is introduced 
in Eq. \eqref{definition of omega tilde}.
Since the integrand function shows two different behaviors according 
to the real or complex nature of $\tilde \omega$ throughout the 
full spectrum, we need a prescription. For $|\omega| >\mu$, we 
have $\tilde \omega=\pm ik$, where $k=\sqrt{\omega^2-\mu^2}$. A 
natural choice describing a purely outgoing wave at infinity 
is given by the retarded branch
\begin{equation}
\tilde \omega_R=\lim_{\varepsilon\rightarrow0^+}
\sqrt{\mu^2-(\omega+i\varepsilon)^2}=-i\text{sgn}(\omega)k,
\end{equation}
while the advanced one is given by
\begin{equation}
\tilde \omega_A(\omega)=\tilde \omega_R(-\omega)
\end{equation}
The symmetrised choice is instead given by 
\begin{equation}
\tilde \omega_S=+ik
\end{equation}
for both signs of $\omega$ and it is investigated in Eqs (51)-(52) 
in Ref. \cite{Esposito:2025jsq} and Eq (90) in Ref. 
\cite{Esposito:2026las}. We choose the retarded one. That choice 
determines in which half-plane we are allowed to close the contour. 
We must be careful in properly analysing the $|\omega|\rightarrow \mu$ 
limit, since $\tilde \omega\rightarrow0$ and therefore not only the 
$B_s$ coefficients diverge, but the $e^{-\frac{b}{\tilde\omega}}$ in 
Eq. \eqref{F integral rewritten fancy} piece exhibits an essential 
singularity, forbidding the integrability of the function.
In general, upon relying on complex analysis, a proper retarded 
definition is to integrate along $\text{Im}(\omega)=c>0$. The following 
results are independent of the specific value of $c$ thanks to the Cauchy theorem.
For such $\omega$ the time exponential reads as $e^{-i\omega t}
=e^{ct}e^{-it\text{Re}(\omega)}$, suggesting a Laplace transform structure. 
Upon setting 
\begin{equation}
p=-i\omega,
\end{equation}
the integration line identified by $\text{Im}(\omega)=c$ becomes 
$\text{Re}(p)=c$. The frequency spectrum $\omega: -\infty\rightarrow\infty$ 
maps to $p:i\infty\rightarrow -i\infty$. Moreover, we define
\begin{equation}
\tilde \omega=p\sqrt{1+\frac{\mu^2}{p^2}},
\end{equation}
 since the principal square root is analytic except where its argument is real and non-positive. $\tilde \omega$ is therefore analytic in $\mathbb{C}-[-i\mu, i\mu]$.
This first change of variable allows us to rewrite the integral 
in Eq. \eqref{F integral rewritten fancy} as the explicit Laplace transform
\begin{align}
F_R(t,r;\mu)=&\int_{c-i\infty}^{c+i\infty}\frac{dp}{i\sqrt{2\pi}}
e^{p t-a\tilde \omega -\frac{b}{\tilde \omega}}\\
\equiv&\sqrt{2\pi}\int_{c-i\infty}^{c+i\infty}\frac{dp}{i2\pi}e^{p t} \Phi(p).
\end{align}
Therefore, the retarded integral ${F_R}/{\sqrt{2\pi}}$ is the 
inverse Laplace transform of $\Phi(p)=e^{-a\tilde \omega
-\frac{b}{\tilde \omega}}$, while the $\text{Re}(p)=c$ line 
is its Bromwich contour.
In terms of $\omega$, the cut is 
therefore given by the real segment $\omega \in [-\mu,\mu]$. The 
only singularity of $\Phi(p)$ is therefore given only by the 
essential singularity in $e^{-b/\tilde \omega}$ when $\tilde \omega$ 
vanishes, i.e., at $p=\pm i\mu$ or $\omega=\mp \mu$. 
There are no further singularities.

Unfortunately, the modulus of $e^{p t} \Phi(p)$ on the Bromwich line, 
$e^{c(t-a)}$ does not decay and therefore it forbids a trivial 
evaluation of the integral. However, we can subtract the nondivergent 
part. Upon expanding $\tilde \omega$ with respect to $p$, we have
\begin{align}
\Phi(p)=&e^{-ap}\left[1-\frac{b+\mu^2a/2}{p}+O(p^{-2})\right]\\
\equiv &e^{-ap}+e^{-ap}g(p).
\end{align}
The first piece is elementary, leading to $\delta(t-a)$. We have
\begin{equation}
F_R=\sqrt{2\pi}\left[\delta(t-a)+\int_{c-i\infty}^{c+i\infty}
\frac{dp}{i2\pi}e^{p(t-a)}g(p)\right].
\end{equation}
We recall that the inverse transform of $e^{-ap}/p$ is 
$\theta(t-a)$, that corresponds to the origin of the tail contribution. 
The last integral converges, provided that $t\neq a$. Indeed, we have
\begin{equation}
    F_R\underset{t\rightarrow a^+}{=}-\sqrt{\frac{\pi}{2}}\mu^2 r.
\end{equation}
This mathematical framework allows us to easily compute the 
integral in the $t<a$ case, where it vanishes since the contour 
contains no singularity. We therefore obtain
\begin{equation}
    F_R=0 \quad t<a.
\end{equation}
This is just causality in Laplace language: no signal reaches the 
radius $r$ before $t=a\sim r_*$ for large $r$. The retarded branch 
indeed places all singularities on or to the left of the imaginary 
$p$-axis. The same argument in the advanced branch holds at $t<a$.

The region denoted by $t>a$ is non-trivial, since the left large 
semicircle contains the cut, including its two essential end points, 
where the square root in $\tilde\omega$ takes opposite values. 
By Cauchy's theorem, we can take any simple contour $\mathcal{C}$ 
that encircles the cut once counter-clockwise and lies in 
$\mathbb{C}-[-i\mu,i\mu]$, having
\begin{equation}
    \frac{F_R}{\sqrt{2\pi}}=\delta(t-a)+\oint_\mathcal{C} \frac{dp}{i2\pi}
e^{pt-a\tilde \omega-\frac{b}{\tilde \omega}}, \quad t>a.
\end{equation}
We just need a well-suited parametrisation. 
We would like to map the $\omega\in[-\mu,\mu]$ interval onto the 
$|w|=1$ circle, and the threshold points $\omega=\pm\mu$ to $w=\pm i$, 
where $w$ is the variable we are looking for. 

A hint is given by the formula $\tilde \omega^2-p^2=\mu^2$, which 
describes a hyperbola where the field mass $\mu$ plays a central 
role and can be easily parametrised by a new variable $w$.
The variable we are looking for is defined by 
\begin{equation}
    w=\frac{1}{\mu}(\tilde \omega+p),
\end{equation}
and the following relations hold:
\begin{align}
    p=&\frac{\mu}{2}(w-w^{-1}),\\
    \tilde\omega=&\frac{\mu}{2}(w+w^{-1}),\\
    dp=&\frac{\tilde \omega}{w}dw.
\end{align}
In this way $p=\infty$ is described by $w=\infty$ with the same 
orientation. Moreover, the cut opens up into the unit circle. 
We notice that, upon setting $w=i\Omega$, $p=\frac{i}{2}\mu 
\left(\Omega+\Omega^{-1}\right)$, which is the classical Joukowski map. 

In light of the above, the contour $\mathcal{C}$ can be taken to 
be any circle $|w|=\rho>1$, running counter-clockwise, leading to
\begin{align}
    \oint_\mathcal{C} \frac{dp}{i2\pi}e^{pt-a\tilde \omega
-\frac{b}{\tilde \omega}}=&\oint_{|w|=\rho}\frac{dw}{i2\pi}
\frac{\tilde \omega(w)}{w}\nonumber\\\times &\text{exp}
\left(\frac{\mu}{2}[(t-a)w-(t+a)w^{-1}]\right.
\nonumber\\ -&\left.\frac{\beta w}{1+w^2}\right),
\label{F integral rewritten not still fancy in the w plane}
\end{align}
where 
\begin{equation}
    \beta=-2M\mu\ln r.
    \label{appendix definition of beta}
\end{equation}

\begin{figure}
    \centering
    \includegraphics[width=1\linewidth]{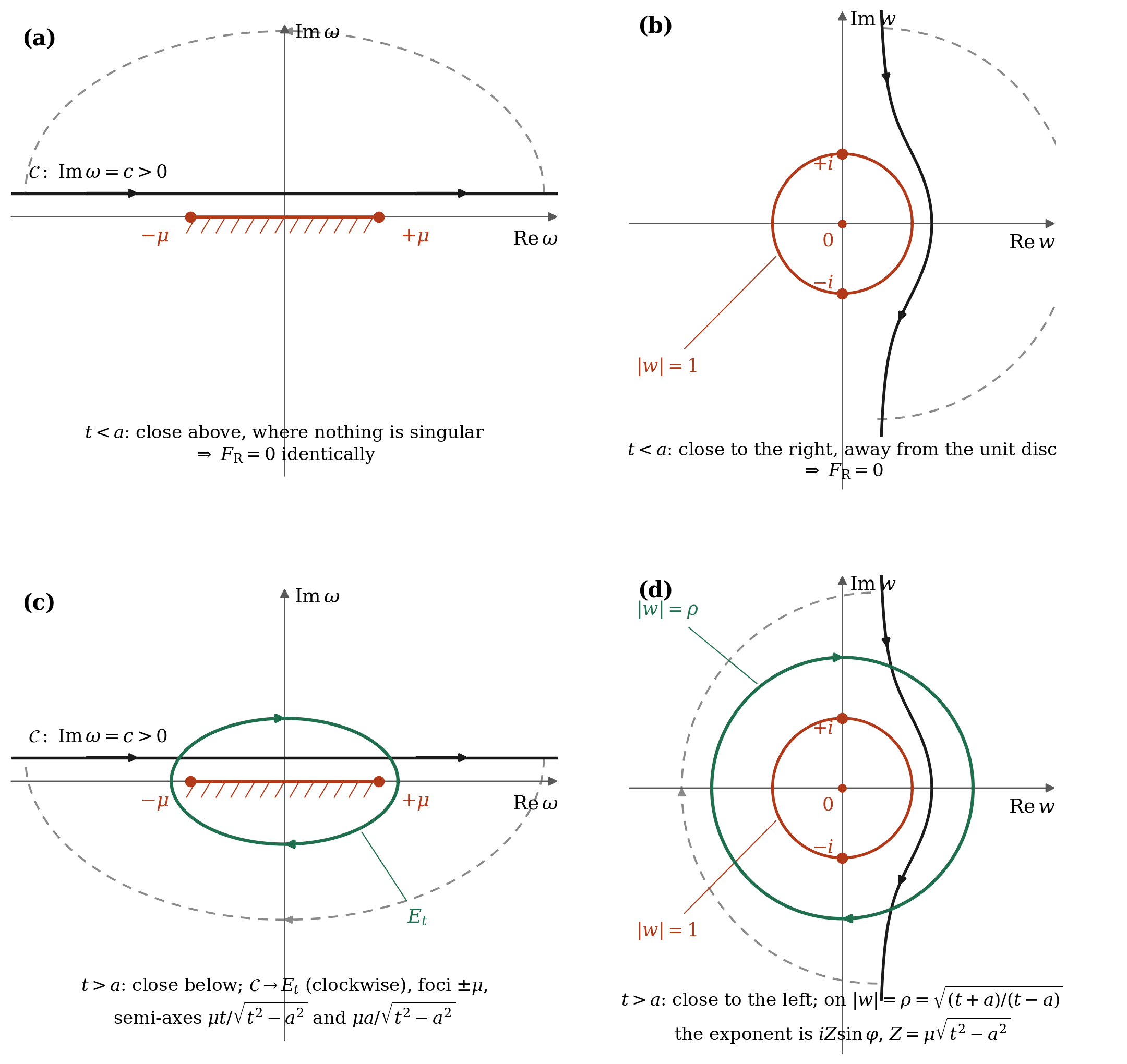}
    \caption{Integration contours for the evaluation of the integral in 
Eq. \eqref{F integral rewritten fancy} in the $\omega$-plane and in 
Eq. \eqref{F integral rewritten not still fancy in the w plane} for the 
$w$-plane. We highlight that the variable transformation $p+\tilde\omega=\mu w$ 
in the $w$ plane maps the ellipse in the $\omega$ plane to the 
$\rho=\sqrt{(t+a)/(t-a)}$, where the exponent is exactly 
$iZ\sin\varphi$. Such a property gives us the possibility to rely on 
Bessel coefficients and an expansion in terms of a Neumann series 
(cf. Sec. ~\ref{appendix section bessel functions}). }
    \label{fig:retarded contour massive scalar in Schwarzschild}
\end{figure}

The integral is now single-valued and it exhibits essential 
singularities only at $w=0$ and $w=\pm i$, lying on or inside the 
unit circle. Hence, the same values are obtained for every $\rho>1$.
These changes of variables and their effect on the integration 
contour are shown in Fig. \eqref{fig:retarded contour massive scalar in Schwarzschild}.

The last and more important step for our purposes is the choice for $\rho$.
The simplest ansatz is
\begin{equation}
    w=\rho e^{i\varphi},
\end{equation}
leading to
\begin{align}
 (t-a)w-(t+a)w^{-1}=&\cos \varphi\left[(t-a)\rho
-\frac{t+a}{\rho}\right]\nonumber\\
 +&i\sin \varphi\left[(t-a)\rho+\frac{t+a}{\rho}\right].
\end{align}
The real part vanishes for every value of the angular variable $\varphi$ provided that
\begin{equation}
    \rho=\sqrt{\frac{t+a}{t-a}}.
\end{equation}
This requires that $t>a$, and then $\rho>1$, hence the circle 
is admissible. Upon defining 
\begin{equation}
    Z=\mu\sqrt{t^2-a^2}
    \label{Appendix definition of Z}
\end{equation}
we have that
\begin{equation}
    \frac{\mu}{2}[(t-a)w-(t+a)w^{-1}]=iZ\sin\varphi
\end{equation}
and then Eq. \eqref{F integral rewritten not still fancy in the w plane} reads as
\begin{align}
    \frac{F_R(t,r)}{\sqrt{2\pi}}=\delta(t-a)+\frac{\theta(t-a)}{2\pi}\int_0^{2\pi}
d\varphi \tilde \omega(\varphi)e^{iZ\sin\varphi 
-\frac{b}{\tilde\omega(\varphi)}},
    \label{FR t>=a}
\end{align}
where
\begin{align}
    \tilde\omega(\varphi)=&\frac{\mu}{\sqrt{t^2-a^2}}
(t\cos\varphi +ia\sin\varphi).
\end{align}
The evaluation of the integral
\begin{equation}
      \mathcal{T}(t,r)=\frac{1}{2\pi}\int_0^{2\pi}d\varphi 
\tilde \omega(\varphi)e^{iZ\sin\varphi -\frac{b}{\tilde\omega(\varphi)}},
\end{equation}
which encodes the tail propagating inside the future-directed 
lightcone (cf. Sec. ~\ref{The Huygens' principle and the decomposition 
of the Green function}), can be performed by noticing the similarity 
with the Bessel integral (cf. Eq. \eqref{appendix bessel 
representation of bessel coefficients}) 
\begin{equation}
    J_\nu(Z)=\frac{1}{2\pi}\int_0^{2\pi}d\varphi e^{i(Z\sin\varphi -\nu\varphi)}.
\end{equation}
The difference relies on the $e^{-\frac{b}{\tilde \omega}}$ piece, which 
can be expanded in a Neumann series (cf. Eq. \eqref{Appendix neumann 
expansion}. Upon defining
\begin{align}
    x=&w^{-1} \quad (|x|<1),\\
   y=&x+\frac{1}{x},
\end{align}
we have
\begin{align}
    \frac{2}{\mu}\tilde \omega e^{-\frac{b}{\tilde \omega}}=&ye^{-\frac{\beta}{y}} \\
    =&\sum_{n\ge0}\frac{(-\beta)^n}{n!}y^{1-n}\\
    =&\sum_{j\ge-1}d_j x^j.
\end{align}
We find that
\begin{equation}
    y^{1-n}=x^{n-1}(1+x^2)^{1-n}.
\end{equation}
This gives the following relations for the $d_j$ coefficients of the series:
\begin{align}
    d_{-1}=&1, \label{d-1 coefficient massive scalar schwarzschild}\\
    d_0=&-\beta, \label{d0 coefficient massive scalar schwarzschild}\\
    d_\nu=&\delta_{\nu1}+\sum_{k=0}^{\frac{\nu-1}{2}}(-1)^k 
\binom{\nu-k-1}{k}\frac{(-\beta)^{\nu+1-2k}}{(\nu+1-2k)!} \quad \nu\ge1. 
\label{dj coefficient massive scalar schwarzschild}
\end{align}
By introducing the auxiliary variable (in Subsec. ~\ref{subsection 
leading order massive scalar field} we rewrite it in terms of retarded 
and advanced time coordinates, since $a\simeq r_*$ in the large-$r$ limit)
\begin{equation}
\xi=\rho^{-1}=\sqrt{\frac{t-a}{t+a}} 
    \label{appendix definition of xi}
\end{equation}
and noting that $J_{-1}=-J_1$ (cf. Eq. \eqref{appendix J-nu=something 
J-nu}), we obtain therefore
\begin{equation}
  \frac{2}{\mu}  \mathcal{T}(t,r)=\sum_{\nu=-1}^\infty d_\nu\xi^\nu J_\nu(Z)
  \label{appendix tail series with dj coefficient massive schwarzschild}
\end{equation}
where $\beta$ and Z are defined in Eqs \eqref{appendix definition of 
beta}, \eqref{Appendix definition of Z} respectively. 
More explicitly, we can write
\begin{align}
\frac{2}{\mu} \mathcal{T}(t,r)=  &-\beta J_0+\left[\left(1
+\tfrac{\beta^2}{2}\right)\xi  -\xi^{-1}\right]  J_1
\nonumber\\&-\tfrac{\beta^3}{6}\xi^2J_2+\big(\tfrac{\beta^4}{24}
-\tfrac{\beta^2}{2}\big)\xi^3J_3\nonumber\\&+\big(\tfrac{\beta^3}{3}
-\tfrac{\beta^5}{120}\big)\xi^4J_4+\dots.
\end{align}
Such a series converges absolutely since $\xi<1$ (this is trivial 
since $\xi$ is the inverse of the radius $\rho$ of the unit circle 
in the complex $w$-plane), the Laurent radius is $1$ and (cf. 
Ref. \cite{NIST:DLMF}, Eq. (10.14.4)) 
\begin{equation}
    |J_\nu(Z)|\le\left(\frac{Z}{2}\right)^\nu\frac{1}{\nu!}.
\end{equation}

The series in Eq. \eqref{appendix tail series with dj coefficient 
massive schwarzschild} can be written in a slightly different 
way. In fact, upon splitting the $d_\nu$ coefficient 
\begin{equation}
    d_\nu=c_{\nu+1}+c_{\nu-1},
\end{equation}
we can elaborate the series in such a way that (we do not rename the 
dummy index $\nu$)
\begin{align}
    \sum_{\nu=-1}^\infty d_\nu\xi^\nu J_\nu
=&\sum_\nu c_\nu\xi^\nu[\xi^{-1}J_{\nu-1}+\xi J_{\nu+1}]
\nonumber\\=&\sum_\nu c_\nu\xi^\nu\left[\frac{t}{\sqrt{t^2-a^2}}(J_{\nu-1}
+J_{\nu+1})\right.\nonumber\\+&\left.\frac{a}{\sqrt{t^2-a^2}}(J_{\nu-1}-J_{\nu+1}) \right]
\end{align}
and, bearing in mind the properties of Bessel functions (cf. 
Refs. \cite{Erdelyi:1953hft, B1945})
\begin{align}
    J_{\nu-1}(Z)+J_{\nu+1}(Z)=&\frac{2 \nu}{Z}J_\nu(Z)\\
    J_{\nu-1}(Z)-J_{\nu+1}(Z)=&2J'_\nu(Z),
\end{align}
and multiplying the result for $\mu/2$, we obtain
\begin{equation}
\mathcal{T}(t,r)=\sum_\nu c_\nu\xi^\nu\left[\frac{\nu t}{t^2-a^2}J_\nu(Z)
+\frac{\mu a}{\sqrt{t^2-a^2}}J'_\nu(Z)\right]
\end{equation}
with the coefficients $c_1=-\beta$, $c_2=\beta^2/2$, $c_3=\beta -\beta^3/6$ defined by
\begin{align}
c_0=&1 \quad \nu=0,\label{Appendix c0 coefficient}\\
c_\nu=&\sum_{k=0}^{(\nu-1)/2}(-1)^k\binom{\nu-k-1}{k}
\frac{(-\beta)^{\nu-2k}}{(\nu-2k)!} \quad \nu\ge1 \label{appendix cnu coefficients}.
\end{align}

This expression allows us to write Eq. \eqref{FR t>=a} in a 
remarkably concise way:
\begin{equation}
    F_R(t,r)=-\sqrt{2\pi}\partial_a\left[\theta(t-a)
\mathcal{J}(t,a;\mu)\right],
\end{equation}
with (we explicitly write the $Z$ variable)
\begin{equation}
    \mathcal{J}(t,a;\mu)=\sum_{\nu=0}^\infty c_\nu(\beta)
\left(\frac{t-a}{t+a}\right)^{\nu/2}J_\nu\left(\mu\sqrt{t^2-a^2}\right).
\end{equation}
This is a generalized Lommel-type series throughout the $c_\nu(\beta)$ 
coefficients (cf Eq. \eqref{appendix Lommel series}).
Some remarkable points. The $\mathcal{J}(t,a;\mu)$ piece reduces 
at $\beta=0$ (i.e., $M=0$) to $J_0\left(\mu\sqrt{t^2-a^2}\right)$, 
which is the Riemann function of the $1+1$ Klein-Gordon propagator. 
The derivative $\partial_a \theta(t-a)$ produces the $\delta(t-a)$ 
front, since only the $c_0J_0(0)=1$ term survives at $t=a$. The 
Schwarzschild correction therefore replaces $J_0$ with the 
$\beta$-deformed Neumann series.  
We underpin once more that Eq. \eqref{F integral rewritten fancy} 
describes the leading term in the large-$r$ expansion of a massive 
scalar field in Schwarzschild spacetime. The result we have obtained 
depends on the relationship among $t$ and the variable $a$ introduced 
in Eq. \eqref{definition of a for the massive scalar schwarzschild 
leading integral}. However, in the large-$r$ limit, we simply find 
that $a\sim r_*$ and Eq. \eqref{FR t>=a} acquires a clear-lightcone 
interpretation, discussed in the main text.

By exploiting the time-reversal symmetry of Schwarzschild spacetime, 
the advanced description is simply obtained by
\begin{equation}
    F_A(t,r)=F_R(-t,r).
\end{equation}
Such a result holds exactly, at any order of the Neumann series. 

\section{Spin-weighted sub-leading terms}

Eq. \eqref{massless field arbitrary spin u r part in utilde} is the truncation at 
$O(r^{-2})$ of the general large-$r$ expansion
\begin{equation}
\frac{\Psi^{(\mathbb{s})}_{\ell 0}(u,r)}{\sqrt{2\pi}}=\sum_{n=0}^{\infty}\frac{h_n(\tilde u)}{r^{n}},
\label{D expansion}
\end{equation}
where we define
\begin{equation}
    h_0(\tilde u)=\delta(\tilde u).
\end{equation}

The $h_n(\tilde u)$ coefficients can be also computed to any order without going through 
the $B^{(\mathbb{s})}_k(\ell,\omega;0)$. The prescription is as follows.
We consider the substitution
\begin{equation}
    \omega\to i\partial_t
\end{equation}
in Eqs \eqref{ps r}, \eqref{qs r}, the change of coordinates from $(t,r)$ to $(\tilde u,r)$ through
\begin{equation}
    \partial_r|_t=\partial_r|_{\tilde u}-f^{-1}\partial_{\tilde u},
\end{equation}
and then we write the radial field as
$r^{-(2\mathbb{s}+1)}\Psi^{(\mathbb{s})}_{\ell0}$. The radial Teukolsky equation therefore 
becomes, for the quantity $X=\mathbf\Psi^{(\mathbb{s})}_{\ell0}/\sqrt{2\pi}$ defined in Eq \eqref{quantity X},
\begin{align}
&-2\,\partial_r\partial_{\tilde u}X+f\,\partial_r^2X
-\left[\frac{2\mathbb{s}}{r}-\frac{2(3\mathbb{s}+1)M}{r^2}\right]\partial_rX
\nonumber\\
&-\left[\frac{\lambda}{r^2}+\frac{2(\mathbb{s}+1)(2\mathbb{s}+1)M}{r^3}\right]X=0 .
\label{D td Teukolsky}
\end{align}
Inserting Eq.~\eqref{D expansion} and collecting the coefficient of $r^{-n-1}$ gives the 
transport recursion for the $h_n(\tilde u)$ coefficients
\begin{align}
2n\,h_n'(\tilde u)={}&\big[\lambda-(n-1)(n+2\mathbb{s})\big]\,h_{n-1}(\tilde u)
\nonumber\\
&+2M(n+\mathbb{s}-1)(n+2\mathbb{s}-1)\,h_{n-2}(\tilde u),
\label{D recursion}
\end{align}
where the prime denotes a total derivative, $h_{-1}\equiv0$, and every step is integrated 
with the retarded condition $h_n(\tilde u)=0$ for $\tilde u<0$. 
For $n=1,2$ it reproduces Eq \eqref{massless field arbitrary spin u r part in utilde}. 
Upon introducing the auxiliary variable, an index-based generalisation of the $\lambda$ in 
Eq. \eqref{lambda spin weighted separation constant},
\begin{align}
\lambda_k\equiv&(\ell-\mathbb{s}-k)(\ell+\mathbb{s}+k+1)\nonumber\\=&\lambda-k(k+2\mathbb{s}+1),
\label{D lambda k}
\end{align}
so that $\lambda_0=\lambda$, the next four orders read
\begin{widetext}
\begin{align}
h_3={}&\frac{\lambda_0\lambda_1\lambda_2}{96}\,\tilde u^2\theta(\tilde u)
+\frac{M}{12}(\mathbb{s}+1)(2\mathbb{s}+3)(3\lambda-4\mathbb{s}-2)\,\tilde u\,\theta(\tilde u),
\label{D h3}\\[4pt]
h_4={}&\frac{\lambda_0\lambda_1\lambda_2\lambda_3}{2304}\,\tilde u^3\theta(\tilde u)
+\frac{M}{96}(\mathbb{s}+2)(2\mathbb{s}+3)\big[3\lambda^2-14(\mathbb{s}+1)\lambda
+6(\mathbb{s}+1)(2\mathbb{s}+1)\big]\tilde u^2\theta(\tilde u)
\nonumber\\
&+\frac{M^2}{8}(\mathbb{s}+1)(\mathbb{s}+3)(2\mathbb{s}+1)(2\mathbb{s}+3)\,\tilde u\,\theta(\tilde u),
\label{D h4}\\[4pt]
h_5={}&\frac{\lambda_0\lambda_1\lambda_2\lambda_3\lambda_4}{92160}\,\tilde u^4\theta(\tilde u)
+\frac{M}{2880}(\mathbb{s}+2)(2\mathbb{s}+5)\big[5\lambda^3-10(5\mathbb{s}+7)
\lambda^2+70(\mathbb{s}+1)(2\mathbb{s}+3)\lambda
\nonumber\\
&-24(\mathbb{s}+1)(2\mathbb{s}+1)(2\mathbb{s}+3)\big]\tilde u^3\theta(\tilde u)
+\frac{M^2}{480}(\mathbb{s}+1)(2\mathbb{s}+3)\big[3(10\mathbb{s}^2+55\mathbb{s}+67)\lambda
\nonumber\\
&-4(2\mathbb{s}+1)(2\mathbb{s}+7)(5\mathbb{s}+11)\big]\tilde u^2\theta(\tilde u),
\label{D h5}\\[4pt]
h_6={}&\frac{\lambda_0\lambda_1\lambda_2\lambda_3\lambda_4\lambda_5}{5529600}\,\tilde u^5\theta(\tilde u)
+\frac{M}{276480}(\mathbb{s}+3)(2\mathbb{s}+5)\big[15\lambda^4-20(13\mathbb{s}+23)\lambda^3
+20(75\mathbb{s}^2+248\mathbb{s}+199)\lambda^2
\nonumber\\
&-1616(\mathbb{s}+1)(\mathbb{s}+2)(2\mathbb{s}+3)\lambda+480(\mathbb{s}+1)(\mathbb{s}
+2)(2\mathbb{s}+1)(2\mathbb{s}+3)\big]\tilde u^4\theta(\tilde u)
\nonumber\\
&+\frac{M^2}{17280}(2\mathbb{s}+3)\big[9(10\mathbb{s}^3+85\mathbb{s}^2+224\mathbb{s}+189)\lambda^2
-6(\mathbb{s}+1)(5\mathbb{s}+13)(22\mathbb{s}^2+133\mathbb{s}+171)\lambda
\nonumber\\
&+20(\mathbb{s}+1)(\mathbb{s}+4)(2\mathbb{s}+1)(26\mathbb{s}^2+127\mathbb{s}+153)\big]\tilde u^3\theta(\tilde u)
\nonumber\\
&+\frac{M^3}{96}(\mathbb{s}+1)(\mathbb{s}+3)(\mathbb{s}+5)(2\mathbb{s}+1)(2\mathbb{s}+3)
(2\mathbb{s}+5)\,\tilde u^2\theta(\tilde u).
\label{D h6}
\end{align}
\end{widetext}
The structure follows a general pattern: in $h_n$ the power $M^j$ multiplies $\tilde 
u^{\,n-1-j}\theta(\tilde u)$, with $0\le j\le\lfloor n/2\rfloor$,
and no $\delta(\tilde u)$ or derivative of it appears for $n\ge1$, hence the prompt 
signal is $\delta(\tilde u)$ to all orders.
Two families of coefficients follow in closed form from Eq.~\eqref{D recursion}. 
At $M=0$ (flat-space limit) only the first term acts, so that
\begin{equation}
h_n\big|_{M=0}=\frac{\lambda_0\lambda_1\cdots\lambda_{n-1}}{2^n\,n!\,(n-1)!}\,
\tilde u^{\,n-1}\theta(\tilde u),\qquad n\ge1,
\label{D flat}
\end{equation}
which for $\mathbb{s}=0$ gives
\begin{equation}
    \lambda_0\cdots\lambda_{n-1}=\frac{(\ell+n)!}{(\ell-n)!},
\end{equation}
i.e.\ the coefficients of $h^{(1)}_\ell$, and terminates at $n=\ell-\mathbb{s}$.
The highest power of $M$ is reached only through the second term, so that
\begin{align}
h_{2j}\supset{}&\frac{M^j}{2^j\,j!\,(j-1)!}\prod_{k=1}^{j}(\mathbb{s}+2k-1)(2\mathbb{s}+2k-1)
\nonumber\\
&\times\tilde u^{\,j-1}\theta(\tilde u),\qquad j\ge1,
\label{D top M}
\end{align}
a purely curvature-induced, $\ell$-independent contribution that vanishes for $\mathbb{s}\in\{-1,-\tfrac12\}$ at every order.

\section{The action of superdilations}

In Ref. \cite{Refuto:2026ljz}, one of the authors studied the emergence of superdilations 
in asymptotically flat spacetimes, explicitly deriving them at Schwarzschild future 
null infinity. Upon relying on the geometrical structure of asymptotic conformal Killing 
horizons, introduced in Ref. \cite{deAguiarAlves:2025dce}, the asymptotic conformal symmetry 
group at future null infinity is given by
\begin{equation}
\Xi_{\mathcal{I}^+}=\text{Conf}(S^2)\ltimes
\big( C^\infty(S^2) \ltimes C^\infty(S^2) \big),
\label{asymptotic conformal group Scwharzschild}
\end{equation}
whose infinitesimal generator is given by
\begin{equation}
       \xi=f(x^A)[\alpha(u\partial_u+r\partial_r)+\partial_u]+Y^A(x^B)\partial_A,
       \label{full xi superdilations Schwarzschild}
\end{equation}
where $f(x^A)\in C^\infty(S^2)$, $\alpha$ is a constant and $Y^A(x^B)$ are the conformal 
Killing vector fields on the two-sphere (no confusion should arise with spherical harmonics 
due to different indices). The starting point of the geometrical construction is to solve 
the conformal Killing equation for a Schwarzschild spacetime in Bondi gauge, leading to 
a set of coupled partial differential equations of order $1/r$. The generator of 
superdilations (the extra $C^\infty(S^2)$ factor in Eq. \eqref{asymptotic conformal 
group Scwharzschild}) is therefore given by
\begin{equation}
    \xi_D\equiv \alpha f(x^A)(\tilde u\partial_{\tilde u}+r\partial_r).
\end{equation}
Since the fields we compute are properly defined in the asymptotic region of Schwarzschild 
spacetime, we aim at finding, if it exists, a possible relation with this enlarged group of symmetries. 
In order to increase readability we introduce the lighter notation
\begin{equation}
\mathbf{\Psi}(\tilde u,r)\equiv\frac{  \mathbf{\Psi}^{(\mathbb{s})}_{\ell 0}(\tilde u,r)}{\sqrt{2\pi}}.    
\end{equation}

We start by considering Eq. \eqref{massless field arbitrary spin u r part in utilde} 
at leading order $1/r$, i.e, 
\begin{equation}
 \mathbf{\Psi}(\tilde u,r)=\delta(\tilde u)
+\frac{\lambda}{2r}\theta(\tilde u)+O\left(r^{-2}\right),
\end{equation}
where the spin-dependence is fully encoded in the separation constant $\lambda$. It 
turns out that it is a homogeneous function of degree $-1$, since
\begin{equation}
    \mathbf{\Psi}(h\tilde u,hr)=h^{-1} \mathbf{\Psi}(\tilde u,r).
\end{equation}
We underpin that such a result holds only at leading order: the curvature terms from $O(Mr^{-2})$ on break the scaling. 
The action of superdilations on such a field is simply given by the Lie derivative
\begin{align}
\mathcal{L}_{\xi_D} \mathbf{\Psi}=&\alpha f(x^A)(\tilde u\partial_{\tilde u}+r\partial_r)  
\mathbf{\Psi}\nonumber\\=&-\alpha f(x^A)  \mathbf{\Psi},
\end{align}
where in the last step we exploited the Euler theorem on homogeneous functions. The prefactor 
$z(x^A)\equiv\alpha f(x^A)$  can be related with the local Iyer-Wald superdilations charge $q_\alpha(u,x^A)$
\begin{equation}
z(x^A)=\frac{4\pi}{M}\frac{d}{du}q_\alpha(u,x^A)
\end{equation}
and can be physically understood as an angle-dependent dilation redshift factor, suppressed 
by the black hole mass $M$. More details in Ref. \cite{Refuto:2026ljz}. The remarkable result 
is that the $1/r$ order of the full spectrum of a massless spin-weighted field at Schwarzschild 
null infinity is conformally rescaled along the flux generated by the superdilations vector 
field $\xi_D$, helping us to shed new lights on this particular asymptotic conformal symmetry 
and its related phenomenology.

\end{document}